\documentclass{aa}  

\usepackage{graphicx}

\usepackage{txfonts}

\usepackage[page]{appendix}
\usepackage{natbib}
\usepackage{array}
\usepackage{longtable}

\usepackage{color}

\newcommand{\epk}{HD~80653}
\begin{document}

   \title{An ultra-short period rocky  super-Earth orbiting the G2-star HD~80653 \thanks{Based on observations made with the Italian {\it Telescopio Nazionale Galileo} (TNG) operated by the {\it Fundaci\'on Galileo Galilei} (FGG) of the {\it Istituto Nazionale di Astrofisica} (INAF) at the  {\it Observatorio del Roque de los Muchachos} (La Palma, Canary Islands, Spain).}\textsuperscript{,} \thanks{HARPS-N spectroscopic data are available in electronic form at the CDS via anonymous ftp to cdsarc.u-strasbg.fr (130.79.128.5) or via http://cdsweb.u-strasbg.fr/cgi-bin/qcat?J/A+A/.}} 

   \author{G.~Frustagli
          \inst{1,2}
          \and
          E.~Poretti \inst{1,3}
          \and
          T.~Milbourne \inst{4,5}
          \and
          L.~Malavolta \inst{6}
          \and 
          A.~Mortier \inst{7}
          \and
          Vikash~Singh \inst{6}
          \and
          A.~S.~Bonomo \inst{8}
          \and
          L.~A.~Buchhave \inst{9}
          \and 
          L.~Zeng \inst{5,10}
          \and
          A.~Vanderburg \inst{5}
          \and
          S.~Udry \inst{11}
          \and
          G.~Andreuzzi\inst{3,12}
          \and
          A.~Collier-Cameron \inst{13}
          \and
          R.~Cosentino \inst{3}
          \and 
          M.~Damasso \inst{8}
          \and 
          A.~Ghedina \inst{3}
          \and
          A.~Harutyunyan \inst{3}
          \and
          R.~D.~Haywood \inst{5}
          \and
          D.~W.~Latham \inst{5}
          \and
          M.~L\'opez-Morales \inst{5}
          \and
          V.~Lorenzi \inst{3,14}
          \and
          A.F.~Martinez~Fiorenzano \inst{3}
          \and
          M.~Mayor \inst{11}
          \and 
          G.~Micela \inst{15}
          \and
          E.~Molinari\inst{16}
          \and
          F.~Pepe \inst{11}
          \and
          D.~Phillips \inst{5}
          \and
          K.~Rice \inst{17,18}
          \and
          A.~Sozzetti  \inst{8} 
}

   \institute{INAF - Osservatorio Astronomico di Brera, Via E. Bianchi 46, 23807 Merate, Italy
              \\
              \email{giuseppe.frustagli@inaf.it}
         \and
           Dipartimento di Fisica G. Occhialini, Universit\`a degli Studi di Milano-Bicocca, Piazza della
           Scienza 3, 20126 Milano, Italy
           \and 
           Fundaci\'on Galileo Galilei-INAF, Rambla Jos\'e Ana Fernandez P\'erez 7, 38712 Bre\~na Baja, TF, Spain
           \and
           Department of Physics, Harvard University, 17 Oxford Street, Cambridge MA 02138, USA
           \and
           Center for Astrophysics ${\rm \mid}$ Harvard {\rm \&} Smithsonian, 60 Garden Street, Cambridge, MA 01238, USA
           \and
           INAF - Osservatorio Astrofisico di Catania, Via S. Sofia 78, 95123 Catania, Italy
           \and
           Astrophysics Group, Cavendish Laboratory, University of Cambridge, J.J. Thomson Avenue, Cambridge CB3 0HE, UK
           \and
           INAF - Osservatorio Astrofisico di Torino, Via Osservatorio 20, 10025 Pino Torinese, Italy
           \and
DTU Space, National Space Institute, Technical University of Denmark, Elektrovej 328, 2800 Kgs. Lyngby, Denmark
           \and
           Department of Earth and Planetary Sciences, Harvard University, 20 Oxford Street, Cambridge, MA 02138, USA
           \and
           Observatoire de Gen\`eve,  D\'epartement d'Astronomie de l'Universit\'e de Gen\`eve, 51 ch. des Maillettes, 1290 Versoix, Switzerland
           \and
           INAF - Osservatorio Astronomico di Roma, Via Frascati 33, 00040 Monte Porzio Catone, Italy
            \and
           Centre for Exoplanet Science, SUPA, School of Physics and Astronomy, University of St.~Andrews,  St.~Andrews, KY169SS, UK 
           \and
           Instituto de Astrof\'isica de Canarias, C/V\'ia L\'actea s/n, 38205 La Laguna, Spain
           \and
           INAF - Osservatorio Astronomico di Palermo, Piazza del Parlamento 1,  90124 Palermo, Italy       
           \and
INAF - Osservatorio Astronomico di Cagliari, Via della Scienza 5, 09047 Selargius, Italy       
            \and
          SUPA, Institute for Astronomy, Royal Observatory, University of Edinburgh, Blackford Hill, Edinburgh EH93HJ, UK
           \and
          Centre for Exoplanetary Science, University of Edinburgh, Edinburgh, UK   
  }

   \date{}


\abstract
{Ultra-short period (USP) planets are a class of exoplanets with periods shorter than one day.  The origin of this sub-population of planets is still unclear, with different formation scenarios highly dependent on the composition of the USP planets. 
A better understanding of this class of exoplanets will, therefore, require an increase in the sample of such planets that 
have accurate and precise masses and radii, which also includes estimates of the level of irradiation and 
information about possible companions.  
Here we report a detailed  characterization of a USP planet around the solar-type star \epk$\equiv$EP\,251279430  using the \textit{K2} light curve and 108 precise radial velocities obtained with the HARPS-N spectrograph, installed on the Telescopio Nazionale Galileo.
From the \textit{K2} C16 data, we found one super-Earth planet ($R_{b}=1.613\pm0.071~R_{\oplus}$) transiting the star on a short-period orbit 
($P_{\rm b}=0.719573\pm0.000021$~d). 
From our radial velocity measurements, we constrained the mass of 
\epk~b to $M_{b}=5.60\pm0.43~M_{\oplus}$. 
We also detected a clear long-term trend in the radial velocity data.
We derived the fundamental stellar parameters and determined a radius of $R_{\star}=1.22\pm0.01~R_{\odot}$ and mass 
of $M_{\star}=1.18\pm0.04~M_{\odot}$, suggesting that \epk\, has an age of $2.7\pm1.2$~Gyr. 
The bulk density ($\rho_{b} = 7.4 \pm 1.1$ g cm$^{-3}$) of the planet is consistent with an Earth-like composition of rock and iron with no thick atmosphere. 
Our analysis of the \textit{K2} photometry also suggests hints of a  shallow secondary eclipse with a depth of 8.1$\pm$3.7~ppm. 
Flux variations along the orbital phase are consistent with zero. The most important contribution might come from the day-side thermal emission from the surface of the planet at $T\sim3480$~K.
}


\titlerunning{HD~80653\,b, an ultra-short period exoplanet}
\authorrunning{G. Frustagli et al.}

   \maketitle
%

\section{Introduction}

The discovery that the most common type of exoplanets with a period less than $\sim$100~d 
has a radius whose length is between that of the Earth (1\,$R_{\oplus}$) and that of Neptune ($\sim 4\,R_{\oplus}$) 
\citep{2009A&A...506..303Q,2013Natur.503..377P},
and with masses below 10\,$M_{\oplus}$ \citep{2011arXiv1109.2497M,2012ApJS..201...15H}
is among the most exciting results in the study of their statistical properties \citep{2017AJ....154..109F,2018AJ....156..264F}. 
It appears that the transition from being rocky and terrestrial to having a substantial gaseous atmosphere 
occurs within this size range \citep{2015ApJ...801...41R}. According to recent studies 
\citep{2017AJ....154..109F,2017RNAAS...1a..32Z,2018MNRAS.479.4786V},
 there is a radius gap 
in the exoplanet distribution between 1.5 and 2\,$R_{\oplus}$, as predicted by  \citet{2013ApJ...775..105O} and 
\citet{2013ApJ...776....2L}. Planets with radii less than $\sim 1.5$\,$R_{\oplus}$ tend to be predominantly rocky, while planets that have radii above  2\,$R_{\oplus}$ sustain a substantial gaseous envelope.

Among small radius exoplanets, the so-called ultra-short period (USP) planets  are of particular interest.
These planets orbit with extremely short periods ($P\leq1$~d), are smaller than about 2$R_{\oplus}$ and appear to have compositions similar to that of the Earth \citep{2018NewAR..83...37W}.
There is also evidence that some of them might have iron-rich compositions  
\citep[e.g., ][]{2018NatAs...2..393S}. The origin of this sub-population of planets is still unclear. According to an early hypothesis, USP planets were originally hot Jupiters that underwent strong photo evaporation \citep{2013ApJ...775..105O,2015Natur.522..459E}, ending up with the complete removal of their gaseous envelope and their solid core exposed. The radius gap could then be explained as being due to highly irradiated, close-in planets losing their gaseous atmospheres, while 
planets on longer period orbits, not being strongly irradiated, are able to retain their atmospheres 
\citep{2017ApJ...847...29O,2017AJ....154..109F,2018MNRAS.479.4786V}. 

Another similar hypothesis suggests that
the progenitors of USP planets are not the hot Jupiters, but are instead the so-called mini-Neptunes, that is planets with rocky cores and hydrogen-helium envelopes, with radii typically between 1.7 and 3.9\,$R_{\oplus}$ and masses lower than $\sim 10$\,M$_{\oplus}$ 
\citep{2017AJ....154...60W}. This origin is compatible with the fact that there is an absence of USP planets with radii between 2.2 and 3.8\,$R_{\oplus}$ \citep{2016NatCo...711201L}, the radius valley between 1.5 and 2\,$R_{\oplus}$ in the planets with periods shorter than 100~d \citep{2017AJ....154..109F}, and that
USP planets are typically accompanied by other planets with periods in the range 1-50~d \citep{2014ApJ...787...47S}. 
In addition, there are also alternative hypotheses, such as USP planets starting on more distant orbits and then migrating 
to their current locations \citep[e.g., ][]{2015MNRAS.448.1729R,2017ApJ...842...40L}, or the in situ formation of rocky 
planets on very short-period orbits \citep[e.g., ][]{2013MNRAS.431.3444C}.

Understanding of the origin and composition of USP planets requires precise and accurate measurements of masses and sizes, along 
with the evaluation of the irradiation received and the presence of companions. The problem is that most of the \textit{Kepler} and \textit{K2} USP candidates orbit stars too faint for precise radial velocity (RV) follow-up. 

In this paper, we report on the discovery and characterization of a USP super-Earth orbiting a bright ($V$=9.4\,mag) G2 star, \epk, based on 
\textit{K2} Campaign~16 photometry  and high-precision HARPS-N spectra. 
This candidate was originally identified by \citet{2018AJ....156...22Y}  in the \textit{K2} raw data, with the
additional comment ``somewhat V-shaped" on the light curve of the transit.

Our paper is structured as follows. Section \ref{sec:Observations} describes the data obtained  from both photometric and spectroscopic observations. Stellar properties, including stellar activity indicators, are discussed in Sect.~\ref{ref:stellar_parameters}. We analyze the transit and the secondary eclipse in Sect.~\ref{sec:photometry}. Section \ref{sec:rv} describes the analysis we performed on the RVs. Finally, we discuss our results and conclusions in Sect.~\ref{sec:discussion}.

\section{Observations} \label{sec:Observations}

\subsection{\textit{K2} photometry}\label{sec:K2_photometry}

\begin{figure}
\centering
      \includegraphics[width=\columnwidth]{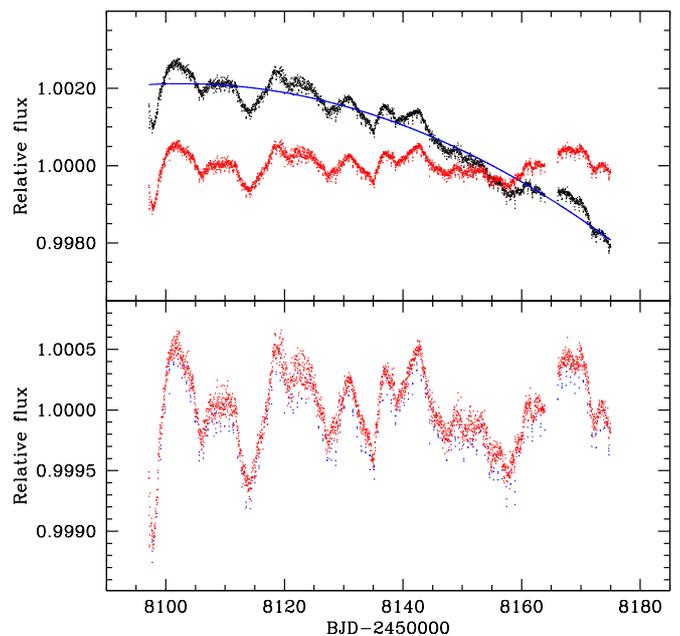}
     \caption{\textit{K2} photometry of \epk. {\it Top panel:} light curve extracted from the MAST raw images (black dots), second-order polynomial fit of the long-term instrumental trend (blue line), corrected light-curve (red dots). {\it Bottom panel:}  corrected data (red dots) with highlighted  measurements obtained during transits (blue dots). 
     Low-frequency flux variations due to the rotational modulation of photospheric active regions are clearly visible.
             }
       \label{Lightcurve}
\end{figure}
\epk\,  was observed by \textit{K2} for about 80 days between 2017 December 9 and 2018 February 25. 
We downloaded the raw images from the Mikulski Archive for Space Telescopes (MAST), 
and we extracted the light curve from the calibrated pixel files following the procedures 
described by \citet{2014PASP..126..948V} and \citet{2016ApJS..222...14V}.
We confirmed Yu et al.'s detection of a planet candidate around the star using 
our Box Least Squares (BLS) transit search pipeline \citep{2002A&A...391..369K,2016ApJS..222...14V}. 
After the identification of the candidate, we re-derived the K2 systematics correction by simultaneously 
modeling the spacecraft roll systematics, planetary transits, and long-term variability \citep{2016ApJS..222...14V}.
The \textit{K2} measurements  show a continuous decrease, very probably due to an instrumental drift 
(black dots in the top panel of Fig.~\ref{Lightcurve}). We removed this drift using a second-order polynomial (blue line), thus obtaining the stellar  photometric behaviour of \epk\, (red dots). 
The preliminary analysis of the \textit{K2} photometry detected the transits, with $P{\rm b}$=0.7195~d and a duration of 1.67~h,
superimposed on a peak-to-valley variability of $\sim 0.1\%$, likely due to the rotational modulation of active regions on the stellar surface (Fig.~\ref{Lightcurve}, bottom panel).

\subsection{HARPS-N spectroscopy}\label{sec:obs_rvs}
The results obtained from the analysis of the \textit{K2} photometry prompted us to include \epk\, in our
HARPS-N Collaboration's Guaranteed Time Observations (GTO) program with the goal of precisely determining the mass of the planet.
We collected 115 spectra from November 2018 to May 2019 with the  HARPS-N spectrograph (R=115\,000) installed on the 3.6-m  Telescopio Nazionale Galileo (TNG), located at the Observatorio del Roque de los Muchachos in La Palma, Spain \citep{2012SPIE.8446E..1VC}.

\begin{figure}
\centering
   \includegraphics[width=\columnwidth]{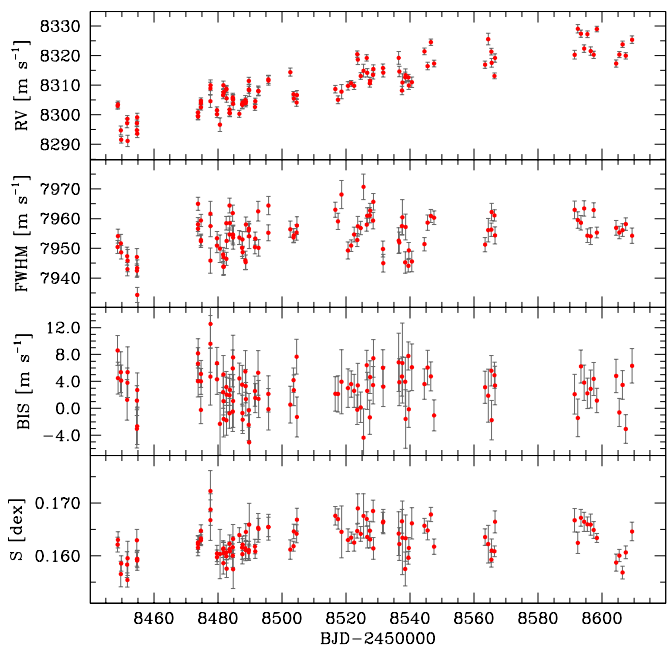}
     \caption{
Radial velocities and stellar activity indicators time series extracted from HARPS- N spectra. A long-term trend is a strong feature in the RV time series, but no counterpart is visible in the indicators. 
}
     \label{Data}
\end{figure}

The spectra were reduced with the version 3.8 of the HARPS-N Data Reduction Software (DRS), which includes corrections for color 
systematics introduced by variations in observing conditions \citep{2014SPIE.9147E..8CC}. A numerical weighted G2 mask was used 
to calculate the weighted cross correlation function \citep[CCF; ][]{2002Msngr.110....9P}. 
The DRS also provides some activity indicators, such as the full width at half maximum (FWHM) of the CCF, the line bisector 
inverse slope (BIS) of the CCF, and the Mount Wilson S-index (S$_{\text{MW}}$). 
We acquired simultaneous Fabry-Perot calibration spectra to correct for the instrumental drift.

All the spectra were taken with an exposure time of 900~s. Due to the short orbital period, we took between 2 and 4 spectra per night on 
several  nights. Seven spectra with  low signal-to-noise ratio (S/N) taken on two nights were no longer considered. The  
108 remaining spectra have S/N in the range 40--127 (median S/N = 84) at 550 nm.

The time series of the RVs and activity indicators are shown in Fig.~\ref{Data}. 
 Error bars on the FWHM and BIS values have been taken as twice those of the RV ones \citep{2018NatAs...2..393S}.
A positive trend is clearly visible for the RVs, with no counterparts
in the stellar activity indicators, pointing out the presence of an outer  companion. We investigate this possibility 
in Sect.~\ref{sec:rv}.

\begin{figure}
\centering
   \includegraphics[width=1.0\columnwidth]{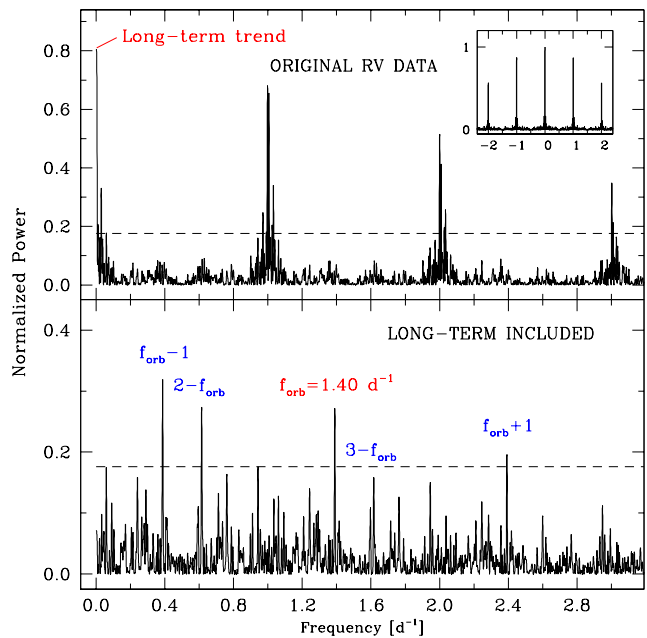}
     \caption{ISW power spectra of the HARPS-N radial velocity data. The horizontal line marks FAP=1\%. 
{\it Top panel:} the long-term trend is the main feature in the original data. The insert shows the spectral window of the data.
{\it Bottom panel:} the orbital frequency of \epk~b is clearly detected when a quadratic trend is added as a known constituent.}
     \label{vani}
\end{figure}

We performed a frequency analysis of the RV time series   using both the 
generalized Lomb-Scargle periodogram \citep[GLS; ][]{2009A&A...496..577Z}. 
and the Iterative Sine-Wave method \citep[ISW; ][]{1971Ap&SS..12...10V}. The latter allowed us to
remove the effects of the prewhitening by recomputing the amplitudes of the 
frequencies and trends  previously identified (indicated as known constituents) 
for each new trial frequency.
As expected, the long-term trend is the most prominent feature (Fig.~\ref{vani}, top panel). Since it is unconstrained
by the time span of the observations, its value is practically 0.0~d$^{-1}$ and the peak structure is very similar to the spectral window (insert in the top panel). The alias structure centered at  the 
orbital frequency $f$=1.40~d$^{-1}$ is already discernible 
in the first power spectrum and  becomes very evident when a quadratic term is introduced in the frequency analysis (bottom panel).
A linear term leaves residual power close to 0.0~d$^{-1}$, evidence  of an unsatisfactory fit. Note that the aliases
are as high as the true peak (see again the spectral window). Indeed, even if we performed more than one measurement per night, the time separation was not very large due to the high number of
nights with the star visible only in the second half of the night. 
This did not allow the effective damping of the $\pm$1~d$^{-1}$ aliases.

\section{Stellar modeling}\label{ref:stellar_parameters}

Table~\ref{table:1a} lists the known stellar parameters of \epk, 
 used in the following analyses. The adopted value for the distance is discussed
in Sect.~\ref{mrad}.

\subsection{Atmospheric parameters}
\label{atmos_param}
We used three different methods to determine the stellar atmospheric parameters. The first method, {\tt CCFpams}\footnote{\url{https://github.com/LucaMalavolta/CCFpams}}, is based on the empirical calibration of temperature, gravity and metallicity on the equivalent width of CCFs obtained with selected subsets of stellar lines, according to their sensitivity to temperature \citep{2017MNRAS.469.3965M}. We obtained $T\mathrm{_{eff}}=5947\pm33$\,K, $\log g = 4.41\pm0.06$~dex (cgs units), and [Fe/H]$=0.30\pm0.03$~dex.

{\tt ARES+MOOG}, the second method we used, is based on the measurement of the equivalent widths of a set of iron lines. For more details, we refer the reader to \cite{2014dapb.book..297S} and references therein. We added all HARPS-N spectra together for this analysis. Equivalent widths were automatically measured using {\tt ARESv2} 
\citep{2015A&A...577A..67S}. The linelist comprises of roughly 300 neutral and ionised iron lines \citep{2011A&A...526A..99S}. Using a grid of ATLAS plane-parallel model atmospheres \citep{1993sssp.book.....K}, the 2017 version of the {\tt MOOG} code\footnote{http://www.as.utexas.edu/~chris/moog.html}
\citep{1973PhDT.......180S}, and assuming local thermodynamic equilibrium, we determined the atmospheric parameters by imposing excitation and ionisation balance. Following the recipe from \citet{2014A&A...572A..95M}, we corrected the surface gravity based on the effective temperature to obtain a more accurate value. Systematic errors were added quadratically to our internal errors \citep{2011A&A...526A..99S}. We obtained $T\mathrm{_{eff}}=6022\pm72$~K, $\log{g}=4.36\pm0.12$~dex and [Fe/H]$=0.35\pm0.05$~dex.

Finally, we used {\tt SPC}, the Stellar Parameter Classification tool \citep{2014Natur.509..593B}, to obtain the atmospheric parameters.
SPC was run on the  individual RV-shifted spectra after which the values of the atmospheric parameters were
averaged and weighted by their S/N. From this method, we obtained
$T_{\rm eff}= 5896\pm50$~K, $\log{g}=4.35\pm0.10$~dex and [Fe/H]$=0.26\pm0.08$~dex.
As {\tt SPC} is a spectral synthesis method,  it also measured the projected rotational velocity, $v\sin i=3.5\pm0.5$~km\,s$^{-1}$.
The $v\sin i$ determinations made with the SPC tool have been shown to be very reliable for such lower rotational velocities
\citep{2012ApJ...757..161T}.

Table~\ref{table:1b} summarizes the parameters determined with the different tools.

%
\begin{table}
\caption{Known stellar parameters of \epk.}
\label{table:1a}      
\centering                          
\begin{tabular}{c  c c }       
\hline\hline                 
\noalign{\smallskip}
  \multicolumn{3}{c}{EPIC 251279430} \\
  \multicolumn{3}{c}{HD 80653} \\
  \multicolumn{3}{c}{2-MASS J09212142+1422046} \\
\hline
\noalign{\smallskip}
\multicolumn{1}{c}{\it Parameter} &\multicolumn{1}{c}{\it Unit}  & \multicolumn{1}{c}{\it Value}\\
\noalign{\smallskip}
\hline
\noalign{\smallskip}
   RA (J2000) & [hms] & 09:21:21.42  \\
   Dec(J2000) & [dms] & 14:22:04.52  \\
\noalign{\smallskip}
    B & [mag] & $10.118 \pm 0.031$ \\ 
    V & [mag] & $9.452 \pm 0.023$ \\ 
    \textit{Kepler} & [mag] & 9.45 \\
    J & [mag] & $8.315 \pm 0.023$ \\ 
    H & [mag] & $8.079 \pm 0.029$ \\ 
    K & [mag] & $8.018 \pm 0.021$ \\     
    W1 & [mag] & $7.959 \pm 0.024$ \\
    W2 & [mag] & $8.000 \pm 0.020$ \\
    W3 & [mag] & $8.011 \pm 0.021$ \\
    W4 & [mag] & $7.869 \pm 0.204$ \\
    Distance & [pc] & $109.86 \pm 0.81$ \\  
\noalign{\smallskip}
\hline\hline
\end{tabular}
\end{table}

\begin{table}
\caption{\epk\, atmospheric parameters. {\tt ARES+MOOG} errors inflated for systematics.
{\tt CCFpams} errors are only internal. {\tt SPC} errors are only internal. }
\label{table:1b}      
\centering                          
\begin{tabular}{c c c c c}       
\hline\hline                 
\noalign{\smallskip}
\multicolumn{1}{c}{Method} & \multicolumn{1}{c}{$T\mathrm{_{eff}}$} & $\log g$ & [Fe/H] & $v \sin i$ \\
\multicolumn{1}{c}{} & \multicolumn{1}{c}{[K]} & \multicolumn{1}{c}{[cgs]}& \multicolumn{1}{c}{[dex]} & \multicolumn{1}{c}{[km s$^{-1}$]} \\
\noalign{\smallskip}
\hline
\noalign{\smallskip}
{\tt ARES+MOOG} & 6022$\pm$72 & 4.36$\pm$0.12 & 0.25$\pm$0.05 & \\
{\tt CCFpams}   & 5947$\pm$33 & 4.41$\pm$0.06 & 0.30$\pm$0.03 & \\
{\tt SPC}       & 5896$\pm$50 & 4.35$\pm$0.10 & 0.26$\pm$0.08 & 3.5$\pm$0.5\\
\noalign{\smallskip}
\hline\hline
\end{tabular}
\end{table}

\subsection{Mass and radius} \label{mrad}

We determined the stellar mass and radius by fitting stellar isochrones using the adopted atmospheric parameters (Sect.~\ref{atmos_param}), 
the apparent B and V magnitudes, photometry from the Two Micron All Sky Survey \citep[2MASS; ][]{2003yCat.2246....0C,2006AJ....131.1163S} 
and the Wide-field Infrared Survey Explorer \citep[WISE: ][]{2010AJ....140.1868W}. 
For the atmospheric parameters, we assumed $\sigma_{T\mathrm{_{eff}}}=70$~K, $\sigma_{\log g}=0.12$~dex and  $\sigma_{\text{[Fe/H]}}=0.08$~dex  as realistic errors for all our parameter estimates, based on the combination of the expected systematic errors \citep[e.g. ][]{2011A&A...526A..99S} and the most conservative internal error estimate for each parameter from all the techniques.

We used the code {\tt isochrones} \citep{2015ascl.soft03010M} to obtain our stellar parameters. The evolutionary models are both the MESA isochrones and Stellar Tracks \citep[MIST; ][]{2011ApJS..192....3P,2016ApJ...823..102C, 2016ApJS..222....8D} and the Dartmouth Stellar Evolution Database \citep{2008ApJS..178...89D}. We ran a fit for each set of stellar atmospheric parameters and repeated the analysis for each set of stellar evolutionary models, for a total of six different fits.

As a final step, we joined the six posterior distributions from the individual fits and calculated the 
median and 16th and 84th percentile of the combined posterior distribution \citep[e.g., ][]{2019MNRAS.484.3731R}. 
We obtained consistent values within errors by using the 
distance values given by the {\it Gaia} DR2 parallax \citep{2016A&A...595A...1G,2018A&A...616A...1G},
by correcting it for a systematic bias \citep{2018ApJ...862...61S} and for 
the nonlinearity of the parallax-distance transformation and the asymmetry of the probability distribution
\citep{2018AJ....156...58B}. 
We used the latter value (Table~\ref{table:1a}), intermediate between the three,  to conclude that 
\epk\, has a  mass 
$M_{\star}=1.18\pm0.04M_{\odot}$, 
a radius of $R_{\star}=1.22\pm0.01\,R_{\odot}$ 
and  an  age of $2.7\pm1.2$~Gyr (Table~\ref{table:1c}).
Note the excellent agreement between the {\it Gaia} and spectroscopic values of $\log~g$ 
(Tables~\ref{table:1b} and \ref{table:1c}).

\begin{table}
\caption{\epk\, stellar parameters from isochrone fits as obtained
from the joined posteriors of six individual fits.
 }
\label{table:1c}      
\centering                          
\begin{tabular}{c c r}       
\hline
\hline
\noalign{\smallskip}
Parameter & Unit & \multicolumn{1}{c}{Value}\\    
\noalign{\smallskip}
\hline
\noalign{\smallskip}
log $g$ & [cgs] &  $4.34\pm0.02$ \\
$M_\star$ & [M$_{\odot}$] & $1.18\pm0.04$ \\
$R_\star$ & [$R_{\odot}$]& $1.22\pm0.01$ \\
Age $t$ & [Gyr]& $2.67\pm1.20$\\
$\log(L_\star/L_{\odot})$ &  & $0.24\pm0.02$\\
$\rho_\star$ & [$\rho_{\odot}$] & $0.64\pm0.04$ \\
\noalign{\smallskip}
\hline \hline
\end{tabular}
\end{table}

\subsection{Stellar activity} \label{activity}

\begin{figure}
\centering
   \includegraphics[width=1.0\columnwidth]{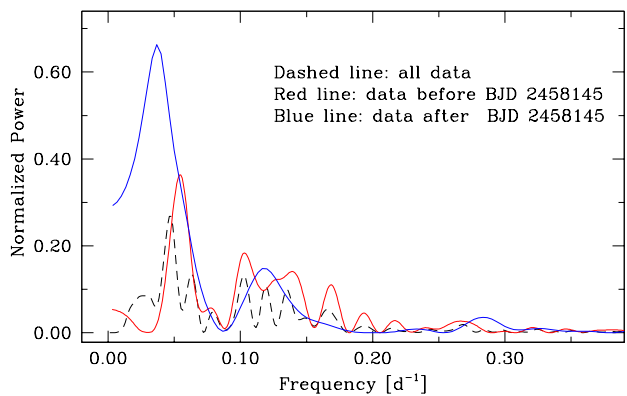}
     \caption{GLS Power spectra of the \textit{K2} data. The full data set has been subdivided into two subsets, before
and after BJD~2458145. Two different rotational frequencies are then detected. 
The power spectrum of the full data set appears as an average of the two.}
       \label{rotation}
\end{figure}

The light curve of \epk\, (Fig.~\ref{Lightcurve}) clearly shows modulated rotational
cycles due to active regions on the stellar surface,
with clearly evident cycle-to-cycle variations. In particular, we note that the standstill at the
level of the average flux  around
BJD~2458145 seems to strongly modify the shape of the light curve. To investigate this,
we firstly removed the in-transit measurements and then
we performed the frequency analysis on the whole dataset and then on  two subsets: the first 
composed of the measurements before BJD~2458145 and the  second composed of those after BJD~2458145.
The resulting GLS power spectra are shown in Fig~\ref{rotation}. They suggest a rotational modulation with 
$f_{\rm rot}$=0.055~d$^{-1}$ and harmonics in the first subset and 
$f_{\rm rot}$=0.037~d$^{-1}$ and harmonics in the second subset. 
The presence of harmonics indicates  a double-wave shape over the
rotational period. This  suggests that the star is seen nearly equator-on with
activity in both hemispheres.
The frequencies  correspond to $P_{\rm rot}$=18~d
and 27~d, respectively, but the latter value is poorly constrained due to the short time coverage.
The power spectrum of the full data set does not supply useful hints, since several
peaks appear as the merging  of the incoherent frequencies detected in the two subsets. 
We can conjecture that in the two time intervals two 
small spots (or groups) appear on well separated regions of the stellar surface.

The measured $v\sin\,i$ (Table~\ref{table:1b}) and inferred stellar radius (Table~\ref{table:1c}) result 
in a maximum rotation period 
$P_{\rm rot}$=18$\pm$3~d. This is in
excellent agreement with the value of the first subset, while that of the second subset  is too long. 
Therefore, it is probable that such a  long period is
spurious due to the simultaneous visibility of several spots widely distributed in longitude.
Taking into account that the flux variability of \epk\, is very small ($\sim$0.1\,\%), the
appearance of small spots can easily alter the light curve. The scenario
becomes still more complicated if the spots are also in differential rotation.

\begin{figure}
\centering
   \includegraphics[width=\columnwidth]{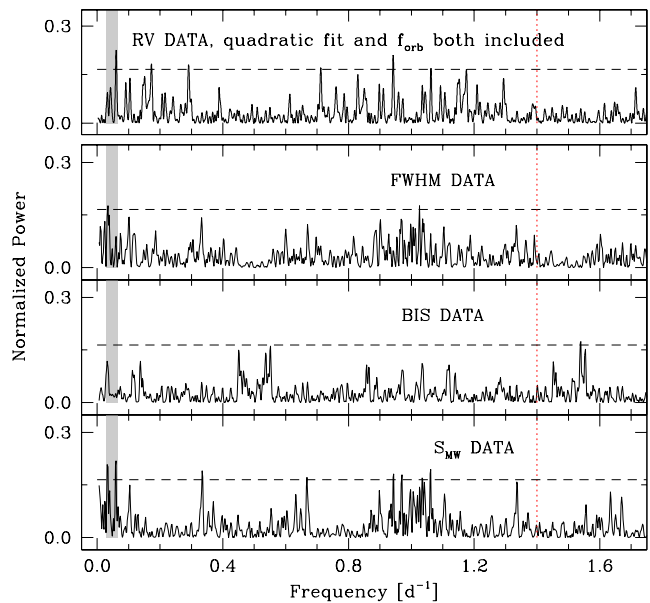}
     \caption{{\it Top to bottom}:  ISW power spectra  of the RV (long-term trend and orbital frequency considered as
known constituents), 
FWHM, BIS, S$_{MW}$ (no known constituent) timeseries. The vertical red line indicates the orbital period of the transiting planet.
The grey region at low frequencies delimits the interval where the rotational frequency is expected from the \textit{K2} photometry.
The horizontal lines mark the 1\% FAP.
             }
       \label{Periodogram}
\end{figure}

We also investigated the periodicities in the spectroscopic time series.
Firstly, we reanalyzed the RV data  
by including the long-term trend and the orbital frequency as known constituents. No clear peak suggesting 
other planetary signals  was detected (Fig.~\ref{Periodogram}, top panel). Then we analyzed the main activity indicators FWHM, BIS, and S$_{MW}$. 
	We immediately noted that none of the periodograms of these indicators show a peak at the frequency $f=1.40~$d$^{-1}$
 detected in the RV data
(Fig.~\ref{Periodogram}, other panels), definitely supporting its full identification as the orbital frequency of \epk~b.
This absence also suggests a weak star-to-planet interaction, if any. On the other hand, these periodograms 
show some peaks in the frequency range where we found signals in the \textit{K2} photometry, namely, 0.03-0.06~d$^{-1}$
(the grey region). In particular, the S$_{MW}$ data and the RV residuals show peaks 
above the FAP=1\% threshold 
at $f$=0.060~d$^{-1}$, that is, $P$=16.6~d. Therefore  \textit{K2} photometry, $v\sin i$ measurement, and activity indicators all suggest
a stellar rotation period in the range 16-20~d.

\begin{figure}
\centering
   \includegraphics[width=\columnwidth]{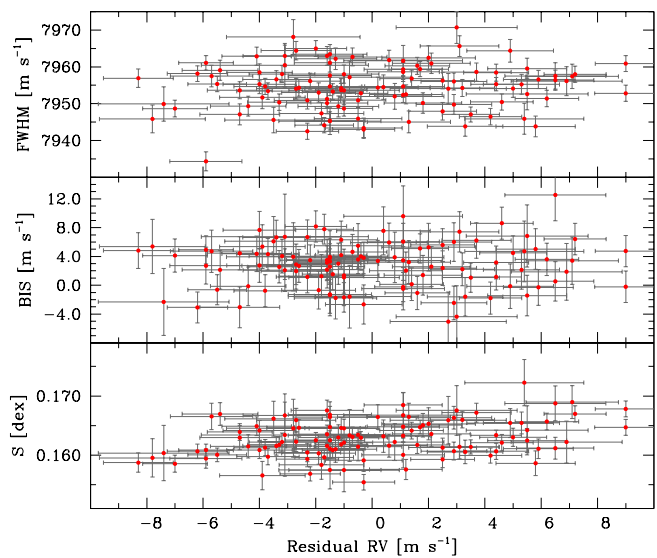}
     \caption{Plots of the RVs versus the activity indicators FWHM, BIS, and S$_{MW}$. 
}
       \label{correl}
\end{figure}

Finally, we calculated the Spearman correlation coefficients for the original RV versus FWHM, BIS, and S$_{MW}$ weighted values 
(Fig.~\ref{correl}).
We obtained $0.355$, $-0.014$, and $0.378$, respectively. 

\section{Photometric modeling} \label{sec:photometry}
The planetary transits are clearly detected in the \textit{K2} light curve, both when extracted
from the raw images (top panel of Fig.~\ref{Lightcurve}) and after correction for the instrumental drift
(bottom panel). This is thanks to  the limited variability ($\sim$0.1\%) of \epk\, and to the sharpness  of the planetary transits.
Moreover, due to the ultra-short orbital period, 
the transits have been observed at almost all the stellar activity levels, thus making very effective the 
cancellation of the effects produced by  unocculted small spots and faculae. 
Therefore, we could perform a 
very reliable analyses both for the transit and the occultation of the exoplanet.

\subsection{Primary Transit}
\label{sec:primary_transit}

We performed the transit fit using \texttt{PyORBIT} 
\citep{2016A&A...588A.118M,2018AJ....155..107M}. We assumed a circular orbit for the planet, applying a parametrisation for the limb darkening \citep{2013MNRAS.435.2152K} 
 and imposing a prior on the stellar density directly derived from the posterior distributions of $M_\star$ and $R_\star$.
We note that a Keplerian fit with no assumption on the eccentricity returned a value consistent with zero.

We removed stellar variability from the \textit{K2} light curve by dividing away 
the best-fit spline from our simultaneous systematics fit described in Sect.~\ref{sec:K2_photometry}, 
and used this flattened light curve in our transit-fit analysis.
\texttt{PyORBIT} relies on
the \texttt{batman} code \citep{2015PASP..127.1161K} to model the transit, with an oversampling factor of 10 
when accounting for the 1764.944~s exposure time of the \textit{K2} observations  \citep{2010MNRAS.408.1758K}. Posterior sampling was performed with an affine-invariant Markov chain Monte Carlo {\it emcee} sampler \citep{2013PASP..125..306F}, with starting points for the chains obtained from
 the global optimization code {\tt PyDE}\footnote{\url{https://github.com/hpparvi/PyDE}}. We ran the sampler for 50000 steps, discarding the first 15000 steps as a conservative burn-in. 

\begin{figure}
\centering
   \includegraphics[width= \columnwidth]{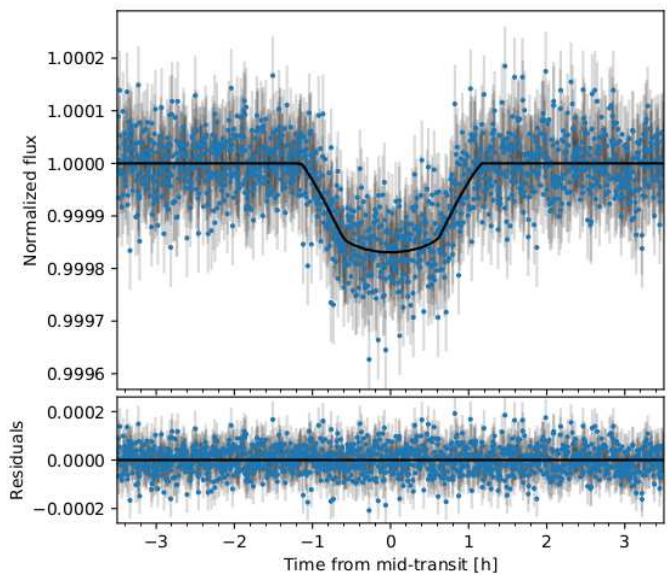}
     \caption{\emph{Top:} \epk~b transit light curve phase-folded to a period of $P_{\rm b}=0.719573$~d, as determined using \texttt{PyORBIT}.  \emph{Bottom}: The residuals of the transit fit. }
             
       \label{Transit}
\end{figure}

We obtained an orbital period $P_{\rm b}=0.719573\pm0.000021$~d 
and a reference central time of transit $T_c=2458134.4244\pm0.0007$~BJD. 
The stellar density as derived from the transit fit agrees with that determined in Sect.~\ref{mrad} (Table~\ref{table:1c}).
The planetary radius is therefore  $R_{\rm b}=1.613\pm0.071$~$R_\oplus$,
given the stellar radius presented in  Sect.~\ref{ref:stellar_parameters}.
We also ran a fit using the raw light curve and modelling it with a Gaussian Process through the {\tt celerite} package \citep{2017AJ....154..220F}. The results were  perfectly consistent with the results from the pre-flattened light curve. All parameters are reported in Table~\ref{table:2} and the phase-folded light curve with the best fit is shown in Fig.~\ref{Transit}. 
 We also measured a much lower value of the \textit{K2} correlated noise (14~ppm) than that
of the photometric errors (60~ppm).
The procedures to compute them are described in
\citet{2006MNRAS.373..231P} and \citet{2012A&A...547A.110B}, respectively.
The resulting total noise is then 62~ppm.

 The values of $v\sin i$, transit depth and impact parameter provide an expected semi-amplitude of 47$\pm$10~cm\,s$^{-1}$
for the Rossiter-McLaughlin effect. This amplitude is smaller than our  $\sigma_{RV}$ errors.

\begin{table}
\caption{\epk\, system parameters.
}             
\label{table:2}      
\resizebox{\columnwidth}{!}{
\begin{tabular}{l l}       
\hline \hline           
\multicolumn{2}{c}{\textit{Fitted parameters}}\\
\hline 
\noalign{\smallskip}
Transit epoch $T_{c}$[BJD$_{TDB}$] & $ 2458134.4244 \pm 0.0007$\\  
\hspace{1.0truecm}  Prior $\mathcal{U}$ [2458134.324, 24588134.524] \\
Orbital Period $P{\rm b}$ [d] & $ 0.719573 \pm 0.000021$ \\ 
\hspace{1.0truecm} Prior $\mathcal{U}$ [0.6195124, 0.8195124] \\
\multicolumn{2}{c}{\textit{Light curve}}\\
LC jitter (normalized flux) & $ 0.000055 \pm 0.000001 $\\
\hspace{1.0truecm} Prior $\mathcal{U}$ [0.000,  0.002]\\ 
{\it Kepler} limb-darkening coefficient $q_{1}$ & $0.31_{-0.19}^{+0.35}$\\
\hspace{1.0truecm} Prior  $\mathcal{U}$ [0.000, 1.000]\\
{\it Kepler} limb-darkening coefficient $q_{2}$ & $0.40_{-0.26}^{+0.33}$\\
\hspace{1.0truecm} Prior  $\mathcal{U}$ [0.000, 1.000]\\
Stellar density $\rho_\star$ [$\rho_\odot$] & 0.64 $\pm$ 0.04\\
\hspace{1.0truecm}  Prior $\mathcal{G}$ [0.64, 0.04]\\
Radius ratio $R_{p}/R_{\star}$ & $ 0.0121 \pm 0.0004$\\
\hspace{1.0truecm} Prior  $\mathcal{U}$ [0.000010, 0.500000]\\
Impact parameter $b$ & $ 0.40_{-0.14}^{+0.08}$ \\
\hspace{1.0truecm} Prior  $\mathcal{U}$ [0.000, 1.000]\\
\multicolumn{2}{c}{\textit{Radial velocities}}\\
RV jitter [m\,s$^{-1}$] & $0.62_{-0.35}^{+0.34}$ \\   
\hspace{1.0truecm}  Prior $\mathcal{U}$ [0.009, 297] \\
Systemic RV $\gamma$ [m\,s$^{-1}$]& $8310.17 \pm 2.2$\\  
\hspace{1.0truecm}  Prior $\mathcal{U}$ [7291.19, 8429.05]\\
Linear term $\dot{\gamma}$ [m\,s$^{-1}$d$^{-1}$]& $0.17 \pm 0.03$\\ 
\hspace{1.0truecm}  Prior $\mathcal{U}$ [-1,+1]\\
Radial-velocity semi-amplitude $K$ [m\,s$^{-1}$] & $3.55 \pm 0.26$\\
\hspace{1.0truecm}  Prior $\mathcal{U}$ in $\log_2$ [-6.64,9.97]\\
Rotational period of the star $P_{\rm rot}$ [d] & $ 19.8 \pm 0.5 $\\ 
\hspace{1.0truecm}  Prior $\mathcal{U}$ [5,40]\\
Coherence scale $w$ &  $ 0.340 \pm 0.034$ \\   
\hspace{1.0truecm}  Prior $\mathcal{G}$ [0.350, 0.035]\\
Decay timescale of activity regions $\lambda$ [d] & $22.5_{-4.7}^{+5.9}$\\  
\hspace{1.0truecm}  Prior $\mathcal{U}$ [5,500]\\
Amplitude of GP $h$ & $5.00_{-0.87}^{+1.22}$\\   
\hspace{1.0truecm}  Prior $\mathcal{U}$  [0.01,100]\\
\noalign{\smallskip}
\hline
\multicolumn{2}{c}{\textit{Derived parameters}}\\
\hline                                   
\noalign{\smallskip}
{\it Kepler} limb-darkening coefficient $u_{1}$ & $0.41_{-0.25}^{+0.34}$\\
{\it Kepler} limb-darkening coefficient $u_{2}$ & $0.10_{-0.30}^{+0.39}$\\
Transit duration $T_{14}$ [d] & $0.0749_{-0.0025}^{+0.0027}$\\
Inclination $i$ [deg] & $82.1\pm2.4$\\
Scaled semi-major axis $a/R_{\star}$ & $2.92\pm0.05$ \\
Orbital semi-major axis $a$ [AU] & $0.0166\pm0.0003$\\

Planet mass $M_{\rm b}$ [M$_{\oplus}$] & $5.60\pm0.43$\\
Planet radius $R_{\rm b}$ [$R_{\oplus}$] & $ 1.613\pm0.071$\\
Planet density $\rho_{\rm b}$ [g cm$^{-3}$] & $7.4\pm1.1 $\\
\noalign{\smallskip}
\hline
\hline
\end{tabular}
}
\end{table}

\subsection{Secondary Eclipse and Phase Curve}

For a few USP planets, namely Kepler-10b \citep{2011ApJ...729...27B}, Kepler-78b \citep{2013ApJ...774...54S} and K2-141b \citep{2018AJ....155..107M}, \textit{Kepler} data have also allowed us to detect the optical secondary eclipse 
and the flux variations along the orbital phase
\footnote{
Exoplanet papers usually call these ``phase variations". However, that is misleading with regards to 
the originary definition used in the study of variable stars. In the latter, ``phase variations" 
indicate variations in the phase values of periodic light-curves, e.g. 
$\phi_i$ in $m(t)=m_o+\sum_i{A_i\cos[2\pi~i~f(t-T_o)+\phi_i]}$,
along the time. 
It thus defines cycle-to-cycle variations, which is not what is meant in the context of exoplanets.}.   
Assuming that the secondary eclipse is mainly due to the planet's thermal emission in the \textit{Kepler} bandpass for the high day-side temperature, the comparison between the depth of the secondary eclipse, $\delta_{\rm ec}$, and the amplitude of the flux variations along the phase, $A_{\rm ill}$, may provide some constraints on the nature of the USP planets. 

To search for the secondary eclipse and flux variations along the phase in \epk, we removed the stellar variability in the \textit{K2} flux curve using the same method as in \citet{2013ApJ...774...54S}. In the resulting filtered flux curve, we simultaneously modeled the primary transit, the secondary eclipse,  and the flux variations along the phase 
by assuming a circular orbit and by using the modified model of \citet{2002ApJ...580L.171M} without limb darkening for the secondary eclipse, and the prescriptions in \citet{2013ApJ...772...51E} for the  flux variations along the phase. Doppler boosting and ellipsoidal variations are negligible and hence were not incorporated in our model. The model was created with a 1-min time sampling and then binned to the long cadence sampling of the \textit{K2} data points. We imposed a Gaussian prior on the stellar density 
(Table~\ref{table:1c})
and fixed the quadratic limb-darkening coefficients to the values previously found (Sect.~\ref{sec:primary_transit} and Table~\ref{table:2}). We employed a differential evolution Markov chain Monte Carlo technique 
 \citep[DE-MCMC; ][]{2006S&C....16..239T} as implemented in the ExofastV2 code \citep{2013PASP..125...83E, 2017ascl.soft10003E} to derive the posterior distributions of the fitted parameters (Fig.~\ref{fig:cornerPlot}).

\begin{figure}
    \centering
    \includegraphics[width=1.10\columnwidth]{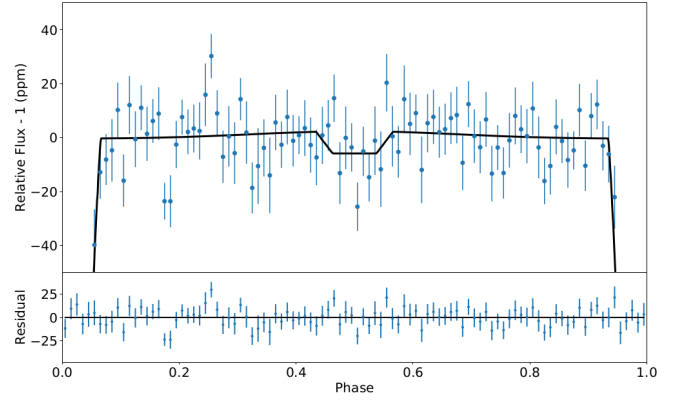}
    \caption{Phase-folded secondary eclipse with the best-fit model and the residuals at the bottom. The data has been binned by a factor of 100 for clarity.}
    \label{fig:SE_PV}
\end{figure}

The values and uncertainties of the primary transit parameters are fully consistent with those determined in Sect.~\ref{sec:primary_transit}.
Our procedure pointed out a possible 
  secondary eclipse with a depth of $\delta_{\rm ec}=8.1\pm 3.7$~ppm at 0.50~$P_{\rm b}$ (Fig.~\ref{fig:SE_PV}).
We computed the values of the Bayesian Information Criterion
\citep[BIC; ][]{2007MNRAS.377L..74L} for the two models with and without the secondary eclipse.  We obtained
$\Delta$BIC=5.2 and hence a Bayes factor $B_{10}\sim13.5$ in favor of the former
\citep{BurnhamAnderson}. By using  appropriate guidelines \citep[see Sect.~3.2 in  ][]{KassRaftery}, this value provides
a positive evidence for the detection of the secondary eclipse. Due to the  
large relative uncertainty on the occultation 
depth, the threshold for a strong detection ($B_{10}=20$) could not  be reached.

\begin{figure}
    \centering
    \includegraphics[width=\columnwidth]{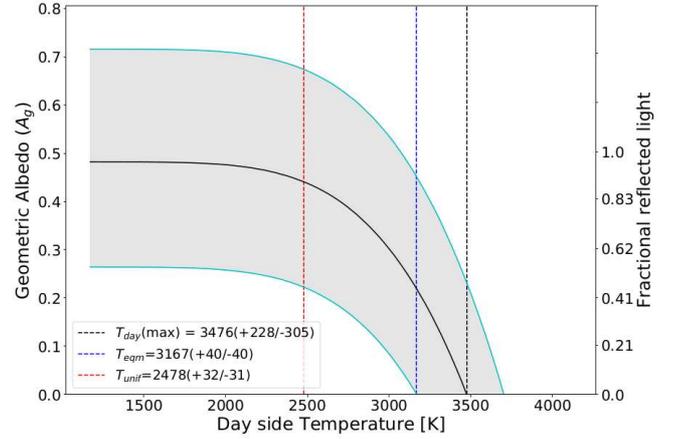}
    \caption{Geometric albedo vs day-side temperature. The dashed lines correspond to the maximum planet day-side temperature (black), the equilibrium temperature in the no-albedo and no-circulation limit (blue) and the uniform temperature for a null Bond albedo and extreme heat circulation (red). 
The shaded grey region displays the 1-$\sigma$ interval for the geometric albedo.}
    \label{fig:albedo}
\end{figure}

The flux variations along the phase are not detectable since they have a predicted null amplitude: $A_{\rm ill}=2.7\pm3.5$~ppm. 
Despite the large uncertainty on $\delta_{\rm ec}$, we estimated the planet geometric albedo as a function of the day-side 
temperature (Fig.~\ref{fig:albedo}). Since the planet is highly irradiated, the most important contribution to the secondary 
eclipse depth might come from the day-side thermal emission rather than the light reflected by the planet surface. 
Figure~\ref{fig:albedo} shows that  
the maximum achievable day-side temperature could be $T_{\rm day}(max)=3476_{-305}^{+228}$~K for a null Bond albedo. 
Theoretical computations for a null Bond albedo and an efficient heat circulation predict a maximum night-side temperature 
$T_{\rm unif}=2478_{-31}^{+32}$~K. 


\begin{figure}
\centering
   \includegraphics[width=\columnwidth]{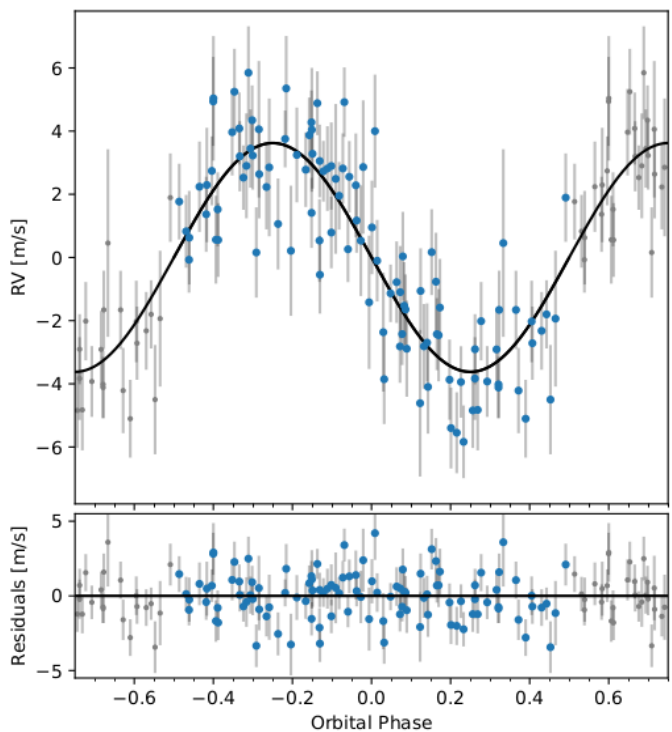}
     \caption{{\it Top panel:} phase-folded RV fit obtained using  {\tt PyORBIT}.
Activity jitter and long-term trend both removed. Larger blue dots delimitate the RV variation over a single orbital cycle.
{\it Bottom panel:} residuals from the best fit.}
       \label{RVcurve2}
\end{figure}

\section{Radial Velocity modeling} \label{sec:rv}

We used two different methods to model the RVs.   
In both methods we assumed a circular orbit. 
This assumption is mainly based on the extremely short circularization time for such a close-in planet, that is, $< 0.5$~Myr
\citep{2008ApJ...686L..29M},
computed from a modified tidal quality factor  similar to that of the Earth ($Q^{'}_{p}\sim1500$),
which is reasonable given  the rocky composition of \epk~b (Table~\ref{table:2} and Sect.~\ref{sec:discussion}).

In the first method we computed the planetary RVs relative to nightly offsets. This was possible
since the orbital period is much shorter than the rotational period. 
We modeled the planetary signal 
fitting for nightly offset values 
calculated every two orbital periods,  using the formula

$$RV = K\sin{\left(\frac{2\pi}{P} x(t)\right)} + \sum_{i=0}^{N}B_i\Theta(t-t_0 - 2iP),$$

\noindent where $P$ is the orbital period determined from the light curve, $\Theta(t)$ is the Heaviside step function, $t_0$ is the time of the first HARPS-N RV measurement, $N$ is the number of nightly offsets, and $x(t)$ is the phase-folded mid-exposure time. 
We find the best-fit parameters via simple likelihood estimate (i.e., $\chi^2$), and then use an MCMC technique 
\citep[as implemented in python by {\it emcee}; ][]{2013PASP..125..306F} to determine appropriate errors on the RV 
amplitude and RV offsets, $K$ and $\left\{ B_i \right\}$ respectively, while $P$ and $T_c$ are determined from the photometric analysis and 
are thus held constant.
 We use flat, non-informative priors for all values and 700 walkers taking 600 steps.
The walkers are initially placed in a small sphere (or Gaussian ball) 
around the values of the Maximum Likelihood Estimate parameters.
The first 50 samples are removed as burn-in, and we then marginalize over all $B_i$s to obtain the posterior distribution for $K$;
 the width of this distribution gives us the errors on the RV amplitude.
We obtained an RV semi-amplitude of 
$K=3.46\pm0.27$~m\,s$^{-1}$ due to planet b, corresponding to a mass  
$M_{\rm b}=5.5\pm0.5$\,M$_{\oplus}$, and a density  $\rho_{\rm b}=7.2\pm1.1$~g\,cm$^{-3}$.

In the second method we used {\tt PyORBIT} to simultaneously model an  orbit 
for the planetary signal 
and a polynomial to account for the  
trend in the RV data (Fig. \ref{Data}). We imposed Gaussian priors based on the light-curve fit 
on $P_{\rm orb}$ and $T_c$ (Sect.~\ref{sec:photometry}).
A first attempt at fitting the data without any modelling of the stellar activity resulted in a RV jitter of 2.9$\pm$0.3~m\,s$^{-1}$, clearly indicating the presence of an additional signal in the data. We then included in our analysis a Gaussian Process (GP) 
trained on  both the S$_{MW}$ index and FWHM, since these indicators show hints of rotational modulation (Sect.~\ref{activity}).
We used a quasi-periodic kernel with 
independent amplitudes of covariance function $h$ for each dataset, 
but with the rotational period $P_{rot}$, coherence scale $w$, and decay timescale of activity regions $\lambda$ in common. We used the {\tt george} library \citep{2015ITPAM..38..252A} to implement the mathematical 
definition of the kernel given by \citet{2015ApJ...808..127G}. We constrained $w$
to $\mathcal{G}(0.35,0.03)$ as suggested by \citet{2016AJ....152..204L}, but taking care to recompute the
proposed prior of $\mathcal{G}(0.50,0.05)$ to take into account the different coefficients in the kernel definition. Non-informative, uniform priors with broad intervals were used for all the other parameters. 
Posterior sampling and confidence intervals were obtained following the same procedure described in Sect.~\ref{sec:photometry}. 

We obtained very consistent results from the GPs trained firstly on the S$_{\text{MW}}$ indicator and then on
the FWHM one.
Both match very well the photometric determination of the rotational period, namely, $P_{\rm rot}$=19.2~d. 
We report the confidence intervals relative to the approach using the 
FWHM index (Table~\ref{table:2}). 
We emphasize that the posterior distributions of the planet's parameters were not affected by the exact choices for the GP regression.
Indeed, for sake of completness, we repeated the analysis using the RV data alone, obtaining again similar results, but larger errors.
The GP analysis yielded an RV semi-amplitude of $K=3.55\pm0.26$~m\,s$^{-1}$ for the planet b, corresponding to a mass 
$M_{\rm b}=5.60\pm0.43$\,M$_{\oplus}$ after taking into account the error on the period, stellar mass and the orbital inclination.

The two methods used to model the RVs yield consistent results on  the RV semi-amplitude
and planetary mass. 
The parameters obtained with {\tt PyORBIT} (Table~\ref{table:2}) were used to  continue our analysis.   
The resulting RV values with the best-fit model are shown in Fig.~\ref{RVcurve2}.

For sake of completness, we performed the {\tt PyORBIT} analysis not assuming a circular orbit. It returned an
eccentricity  value consistent with zero and  parameters all consistent with the circular case. 
We also repeated the  analysis described above to search for another short-period planet in the system, in a circular or eccentric orbit, but we did not detect any clear signal.

The long-term trend seen in the RV plot only
(Fig.~\ref{Data}) strongly suggests an additional Keplerian motion, since long-term activity cycles
should show analogues in the specific indicators and HARPS-N has a proven long-term instrumental stability.   
The steady increase ($\dot{\gamma}$=0.17~m\,s$^{-1}$d$^{-1}$) spans about 150 days. 
There is a relation 
 between $\dot{\gamma}$ and some properties of the companion \citep{2009ApJ...703L..99W}

$$    \frac{m_c \sin{i_c}}{a_c^{2}} \sim \frac{\dot{\gamma}}{G} = (0.37 \pm 0.08)\,{\rm M_{Jup} AU}^{-2},$$

where $m_{c}$ is the companion mass, $i_{c}$ its orbital inclination relative to the line of sight, and $a_{c}$ its orbital distance.
Assuming $i\simeq90^{\rm o}$, a substellar companion of 15~M$_{\rm Jup}$ would orbit at 6.4~AU, with a period of  
 5500~d.
However, the last RV measurements seem to suggest a possible curvature, as noted in the frequency analysis (Sect.~\ref{sec:obs_rvs}).
In such a  case, and assuming a moderate eccentricity, we can estimate a period in the range 260$-$400~d and
a RV amplitude $2K_c\simeq$25~m\,s$^{-1}$. Under these hypothetical conditions, we can tentatively  suggest a mass 
of 0.35$-$0.50~M$_{\rm Jup}$ for the companion. 
The only way to solve the ambiguities on its presence and location is to continue monitoring the system in future observing seasons.



\section{Discussion and Conclusions} \label{sec:discussion}

We used \textit{K2} photometry and high-resolution spectroscopy to determine the \epk\, system parameters and, in particular, the mass and density of its USP transiting planet. A combined analysis of the high-precision HARPS-N RVs and the \textit{K2} data reveals that this planet has an orbital period 
$P_{\rm b}=0.719573\pm0.000021$~d, a radius $R_{\rm b}=1.613\pm0.071~R_{\oplus}$, and a mass $M_{\rm b}=5.60\pm0.43~M_{\oplus}$.
Its density is then $\rho_{\rm b}=7.4\pm1.1$~g\,cm$^{-3}$.

   \begin{figure*}
   \centering

 \includegraphics[width=\textwidth]{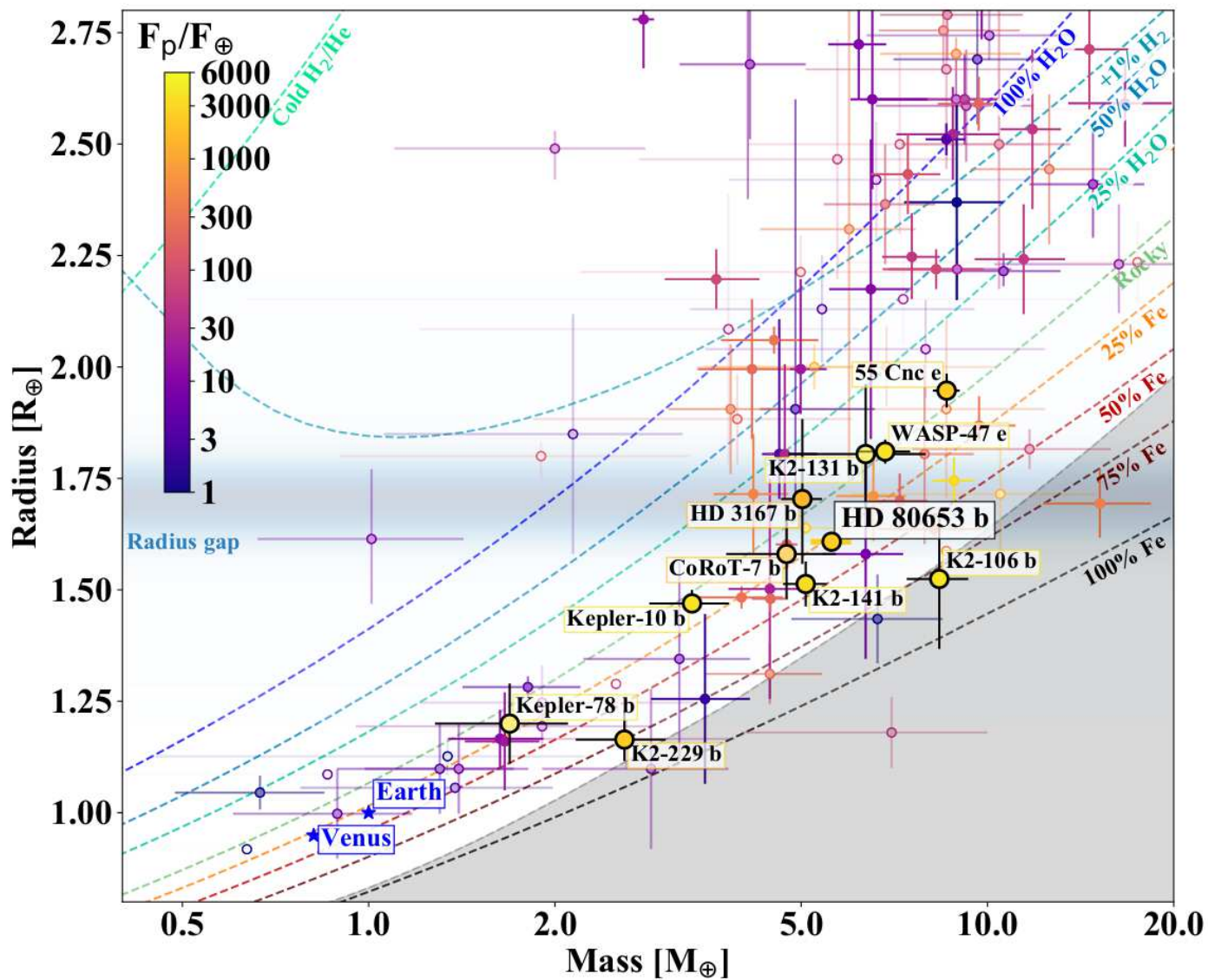}
      \caption{
Mass-radius diagram of planets smaller than $\sim2.8\,R_{\oplus}$.
The data points are shaded according to the precision on the mass, with a full color indicating a value better than 20\%. 
Earth and Venus are shown for comparison. The dashed lines show planetary interior models for different compositions as labelled \citep{2019PNAS.Zeng9723}.  Planets are color-coded according to the incident flux Fp, relative to the solar constant F$_{\odot}$
The horizontal light-blue shade centered on $R\sim1.70\,R_{\oplus}$
shows the radius  Gap. 
The shaded gray region marks the maximum value of iron content predicted by collisional stripping \citep{2010ApJ...712L..73M}.
              }
         \label{RvsM}
         
   \end{figure*}

Figure \ref{RvsM} shows the mass-radius diagram for all small planets ($R_{p}<2.8~R_{\oplus}$) with a 
mass determined with a precision better than $30\%$\footnote{Data from \url{exoplanetarchive.ipac.caltech.edu} (Exoplanet Archive) and
\url{www.exoplanet.eu} \citep{2011A&A...532A..79S}.}. 
\epk~b has a bulk density consistent, within uncertainties,  with that of an Earth-like rocky composition (32.5\% Fe/Ni- metal + 67.5\% Mg-silicates-rock). Owing to its proximity to the host star (1/60~AU),
1Gand the host star being brighter than our own Sun (a factor of 1.67 times the bolometric luminosity of the Sun), \epk~b receives very high bolometric irradiation (radiation flux received per unit surface area is about 6,000 times the insolation at the Earth's surface). Thus, its  equilibrium surface temperature for null albedo and uniform redistribution $T_{\rm unif}$ of heat to the night-side is on the order of 2500~K, enough to completely melt most silicates-rocky materials as well as iron and its alloys under 1-bar surface pressure. Therefore, it is expected to be a lava-ocean world, especially at the substellar point facing the star. 
The uncertainty in our mass determination is small enough that our density estimate excludes the presence of any significant envelope of volatiles
or H/He on the surface of the planet.

Other similar exoplanets can provide useful information about the physical conditions  on the surface of \epk~b.
55~Cnc~e is a super-Earth orbiting a sun-like star in 0.7~d \citep{2018ApJ...860..122C}. 
{\it Spitzer} infrared observations 
\citep{2016Natur.532..207D} show that 55~Cnc~e  is tidally-locked to the host star, meaning that one hemisphere of the
planet always faces the host star, with a hot spot phase-shifted eastward of the substellar point by about 40 degrees. Furthermore, the 
4.5~$\mu$m  phase curve shows that there is a significant temperature difference, on the order of $\sim$1000~K, between the day-side and night-side of the planet.   
Under the intense stellar irradiation on the day-side, which results in the estimated high temperature, there is likely a hemispherical silicate-vapor atmosphere developed on top of the molten liquid silicates \citep[magma pool; ][]{2016ApJ...828...80K}. This silicate-vapor atmosphere would include gaseous species such as SiO 
and Na \citep{Schaefer_2009_ApJ_703_L113}. 
On the other hand, LHS~3844~b is  a 1.3~R$_\oplus$\, world orbiting  a small-size, low-mass and cool star 
($R_\star=0.18~R_\odot, M=0.16~M_\odot,$ and $T\mathrm{_{eff}}=3036~K$) in 0.46~d. It has been modeled as a bare-rock planet, with no atmosphere, but
unfortunately we do not know the mass \citep{Kreidberg2019}.
{\it Spitzer} light curve shows symmetric, large amplitude flux variations along the orbital phase, 
implying a day-side $T=1040~K$ and a night-side close to 0~K.

In the case of \epk,  
 assuming that both the depth of the secondary eclipse ($8.1\pm3.7$~ppm) and the flux variations along the phase 
(consistent with zero) are due to the planet's thermal light, this might indicate 
 non-negligible night-side emission, which would be at odds with the lava-ocean planet model \citep{2011Icar..213....1L} 
predicting  inefficient circulation. 
Two reasons prevent us from drawing any firm conclusion: {\it i})  the large uncertainties on the amplitudes of the secondary eclipse and of the flux variations along the phase, {\it ii}) the well-known degeneracy between reflected and thermal light in the \textit{Kepler} optical bandpass \citep{2011ApJ...729...54C}.  Precise space-based photometry 
(e.g., JWST, CHEOPS) would be extremely useful to unveil the nature of the USP super-Earth \epk~b.

We may also consider the process by which \epk~b formed.
It is unlikely that such an USP planet could form in situ, because a simple equilibrium condensation calculation shows that Magnesium-Silicates, one of the major chemical components of rocks, would only condense out of the nebula from gas phase into solid phase below around 1400~K~\citep{Lewis2004}. 
It is more likely that it initially formed on a wider orbit  and was subsequently transported to its current proximity 
to the star through migration.

 In such a scenario the unseen companion could play a relevant role.  
Indeed, \epk\, exhibits a long-term RV trend, with possible hints of curvature towards the end of the observing period. 
This is suggestive of the existence of an outer, 
more massive companion. Continued RV monitoring of the system, significantly extending  the present time baseline, would enable tight
constraints on its orbital parameters and mass, 
thereby allowing investigations of its role in the formation  of the USP 
super-Earth \epk~b. Given that \epk\, is bright, and in proximity of the Sun, intermediate-separation giant planetary and brown dwarf 
companions are likely to be detectable using {\it Gaia} \citep[e.g., ][ and references therein]{2018haex.bookE..81S}. 
In less than two years time, the third major Gaia Data Release, based on about three years of data collection, might allow 
us to place an additional, independent constraint on the orbital architecture of the \epk\, planetary system.

\begin{acknowledgements}
The HARPS-N project has been funded by the Prodex Program of the Swiss Space Office (SSO), the Harvard University Origins of Life Initiative (HUOLI), the Scottish Universities Physics Alliance (SUPA), the University of Geneva, the Smithsonian Astrophysical Observatory (SAO), and the Italian National Astrophysical Institute (INAF), the University of St Andrews, Queen's University Belfast, and the University of Edinburgh.
Based on observations made with the Italian {\it Telescopio Nazionale Galileo} (TNG) operated by the {\it Fundaci\'on Galileo Galilei} (FGG) of the {\it Istituto Nazionale di Astrofisica} (INAF) at the  {\it Observatorio del Roque de los Muchachos} (La Palma, Canary Islands, Spain).  

G.F. acknowlwdges support from the INAF/FRONTIERA project through the ``Progetti Premiali" funding scheme of the Italian
Ministry of Education, University, and Research.
A.M. acknowledges support from the senior Kavli Institute Fellowships.
A.C.C. acknowledges support from the Science and Technology Facilities Council (STFC) consolidated grant number ST/R000824/1 and UKSA grant ST/R003203/1.
F.P. kindly acknowledges the Swiss National Science Foundation for its continuous support to the 
HARPS-N GTO programme through the grants Nr. 184618, 166227 and 200020\_152721.
R.D.H. and A.V performed this work under contract with the California Institute of Technology 
(Caltech)/Jet Propulsion Laboratory (JPL) funded by NASA through the Sagan Fellowship Program executed by the NASA Exoplanet Science Institute.

This research has made use of the SIMBAD database, operated at CDS, Strasbourg, France, and NASA's Astrophysics Data System.
This research has made use of the NASA Exoplanet Archive, which is operated by the California Institute of Technology, under contract with the National Aeronautics and Space Administration under the Exoplanet Exploration Program.
This material is based upon work supported by the National Aeronautics and Space Administration under grant No. NNX17AB59G issued through the Exoplanets Research Program. 
This work has made use of data from the European Space Agency (ESA) mission
{\it Gaia} (\url{https://www.cosmos.esa.int/gaia}), processed by the {\it Gaia}
Data Processing and Analysis Consortium (DPAC, \url{https://www.cosmos.esa.int/web/gaia/dpac/consortium}). Funding for the DPAC
has been provided by national institutions, in particular the institutions
participating in the {\it Gaia} Multilateral Agreement.
This paper includes data collected by the \textit{K2} mission. Funding for the \textit{K2} mission is provided by the NASA Science Mission directorate. Some of the data presented in this paper were obtained from the Mikulski Archive for Space Telescopes (MAST). STScI is operated by the Association of Universities for Research in Astronomy, Inc., under NASA contract NAS5-26555. Support for MAST for non-HST data is provided by the NASA Office of Space Science via grant NNX13AC07G and by other grants and contracts.
\end{acknowledgements}

\bibliographystyle{aa} 
\bibliography{HD80653_Frustagli+}

\begin{thebibliography}{88}
\expandafter\ifx\csname natexlab\endcsname\relax\def\natexlab#1{#1}\fi

\bibitem[{{Ambikasaran} {et~al.}(2015){Ambikasaran}, {Foreman-Mackey},
  {Greengard}, {Hogg}, \& {O'Neil}}]{2015ITPAM..38..252A}
{Ambikasaran}, S., {Foreman-Mackey}, D., {Greengard}, L., {Hogg}, D.~W., \&
  {O'Neil}, M. 2015, IEEE Transactions on Pattern Analysis and Machine
  Intelligence, 38, 252

\bibitem[{{Bailer-Jones} {et~al.}(2018){Bailer-Jones}, {Rybizki}, {Fouesneau},
  {Mantelet}, \& {Andrae}}]{2018AJ....156...58B}
{Bailer-Jones}, C.~A.~L., {Rybizki}, J., {Fouesneau}, M., {Mantelet}, G., \&
  {Andrae}, R. 2018, \aj, 156, 58

\bibitem[{{Batalha} {et~al.}(2011){Batalha}, {Borucki}, {Bryson}, {Buchhave},
  {Caldwell}, {Christensen-Dalsgaard}, {Ciardi}, {Dunham}, {Fressin},
  {Gautier}, {Gilliland}, {Haas}, {Howell}, {Jenkins}, {Kjeldsen}, {Koch},
  {Latham}, {Lissauer}, {Marcy}, {Rowe}, {Sasselov}, {Seager}, {Steffen},
  {Torres}, {Basri}, {Brown}, {Charbonneau}, {Christiansen}, {Clarke},
  {Cochran}, {Dupree}, {Fabrycky}, {Fischer}, {Ford}, {Fortney}, {Girouard},
  {Holman}, {Johnson}, {Isaacson}, {Klaus}, {Machalek}, {Moorehead},
  {Morehead}, {Ragozzine}, {Tenenbaum}, {Twicken}, {Quinn}, {VanCleve},
  {Walkowicz}, {Welsh}, {Devore}, \& {Gould}}]{2011ApJ...729...27B}
{Batalha}, N.~M., {Borucki}, W.~J., {Bryson}, S.~T., {et~al.} 2011, \apj, 729,
  27

\bibitem[{{Bonomo} {et~al.}(2012){Bonomo}, {Chabaud}, {Deleuil}, {Moutou},
  {Bouchy}, {Cabrera}, {Lanza}, {Mazeh}, {Aigrain}, {Alonso}, {Guterman},
  {Santerne}, \& {Schneider}}]{2012A&A...547A.110B}
{Bonomo}, A.~S., {Chabaud}, P.~Y., {Deleuil}, M., {et~al.} 2012, \aap, 547,
  A110

\bibitem[{{Buchhave} {et~al.}(2014){Buchhave}, {Bizzarro}, {Latham},
  {Sasselov}, {Cochran}, {Endl}, {Isaacson}, {Juncher}, \&
  {Marcy}}]{2014Natur.509..593B}
{Buchhave}, L.~A., {Bizzarro}, M., {Latham}, D.~W., {et~al.} 2014, \nat, 509,
  593

\bibitem[{Burnham \& Anderson(2004)}]{BurnhamAnderson}
Burnham, K.~P. \& Anderson, D.~R. 2004, Sociological Methods \& Research, 33,
  261

\bibitem[{{Chiang} \& {Laughlin}(2013)}]{2013MNRAS.431.3444C}
{Chiang}, E. \& {Laughlin}, G. 2013, \mnras, 431, 3444

\bibitem[{{Choi} {et~al.}(2016){Choi}, {Dotter}, {Conroy}, {Cantiello},
  {Paxton}, \& {Johnson}}]{2016ApJ...823..102C}
{Choi}, J., {Dotter}, A., {Conroy}, C., {et~al.} 2016, \apj, 823, 102

\bibitem[{{Cosentino} {et~al.}(2014){Cosentino}, {Lovis}, {Pepe}, {Cameron},
  {Latham}, {Molinari}, {Udry}, {Bezawada}, {Buchschacher}, {Figueira},
  {Fleury}, {Ghedina}, {Glenday}, {Gonzalez}, {Guerra}, {Henry}, {Hughes},
  {Maire}, {Motalebi}, \& {Phillips}}]{2014SPIE.9147E..8CC}
{Cosentino}, R., {Lovis}, C., {Pepe}, F., {et~al.} 2014, in Society of
  Photo-Optical Instrumentation Engineers (SPIE) Conference Series, Vol. 9147,
  Ground-based and Airborne Instrumentation for Astronomy V, 91478C

\bibitem[{{Cosentino} {et~al.}(2012){Cosentino}, {Lovis}, {Pepe}, {Collier
  Cameron}, {Latham}, {Molinari}, {Udry}, {Bezawada}, {Black}, {Born},
  {Buchschacher}, {Charbonneau}, {Figueira}, {Fleury}, {Galli}, {Gallie},
  {Gao}, {Ghedina}, {Gonzalez}, {Gonzalez}, {Guerra}, {Henry}, {Horne},
  {Hughes}, {Kelly}, {Lodi}, {Lunney}, {Maire}, {Mayor}, {Micela}, {Ordway},
  {Peacock}, {Phillips}, {Piotto}, {Pollacco}, {Queloz}, {Rice}, {Riverol},
  {Riverol}, {San Juan}, {Sasselov}, {Segransan}, {Sozzetti}, {Sosnowska},
  {Stobie}, {Szentgyorgyi}, {Vick}, \& {Weber}}]{2012SPIE.8446E..1VC}
{Cosentino}, R., {Lovis}, C., {Pepe}, F., {et~al.} 2012, in Society of
  Photo-Optical Instrumentation Engineers (SPIE) Conference Series, Vol. 8446,
  Ground-based and Airborne Instrumentation for Astronomy IV, 84461V

\bibitem[{{Cowan} \& {Agol}(2011)}]{2011ApJ...729...54C}
{Cowan}, N.~B. \& {Agol}, E. 2011, \apj, 729, 54

\bibitem[{{Crida} {et~al.}(2018){Crida}, {Ligi}, {Dorn}, \&
  {Lebreton}}]{2018ApJ...860..122C}
{Crida}, A., {Ligi}, R., {Dorn}, C., \& {Lebreton}, Y. 2018, \apj, 860, 122

\bibitem[{{Cutri} {et~al.}(2003){Cutri}, {Skrutskie}, {van Dyk}, {Beichman},
  {Carpenter}, {Chester}, {Cambresy}, {Evans}, {Fowler}, {Gizis}, {Howard},
  {Huchra}, {Jarrett}, {Kopan}, {Kirkpatrick}, {Light}, {Marsh}, {McCallon},
  {Schneider}, {Stiening}, {Sykes}, {Weinberg}, {Wheaton}, {Wheelock}, \&
  {Zacarias}}]{2003yCat.2246....0C}
{Cutri}, R.~M., {Skrutskie}, M.~F., {van Dyk}, S., {et~al.} 2003, VizieR Online
  Data Catalog, II/246

\bibitem[{{Demory} {et~al.}(2016){Demory}, {Gillon}, {de Wit}, {Madhusudhan},
  {Bolmont}, {Heng}, {Kataria}, {Lewis}, {Hu}, {Krick}, {Stamenkovi{\'c}},
  {Benneke}, {Kane}, \& {Queloz}}]{2016Natur.532..207D}
{Demory}, B.-O., {Gillon}, M., {de Wit}, J., {et~al.} 2016, \nat, 532, 207

\bibitem[{{Dotter}(2016)}]{2016ApJS..222....8D}
{Dotter}, A. 2016, \apjs, 222, 8

\bibitem[{{Dotter} {et~al.}(2008){Dotter}, {Chaboyer}, {Jevremovi{\'c}},
  {Kostov}, {Baron}, \& {Ferguson}}]{2008ApJS..178...89D}
{Dotter}, A., {Chaboyer}, B., {Jevremovi{\'c}}, D., {et~al.} 2008, \apjs, 178,
  89

\bibitem[{{Eastman}(2017)}]{2017ascl.soft10003E}
{Eastman}, J. 2017, {EXOFASTv2: Generalized publication-quality exoplanet
  modeling code}

\bibitem[{{Eastman} {et~al.}(2013){Eastman}, {Gaudi}, \&
  {Agol}}]{2013PASP..125...83E}
{Eastman}, J., {Gaudi}, B.~S., \& {Agol}, E. 2013, \pasp, 125, 83

\bibitem[{{Ehrenreich} {et~al.}(2015){Ehrenreich}, {Bourrier}, {Wheatley},
  {Lecavelier des Etangs}, {H{\'e}brard}, {Udry}, {Bonfils}, {Delfosse},
  {D{\'e}sert}, {Sing}, \& {Vidal-Madjar}}]{2015Natur.522..459E}
{Ehrenreich}, D., {Bourrier}, V., {Wheatley}, P.~J., {et~al.} 2015, \nat, 522,
  459

\bibitem[{{Esteves} {et~al.}(2013){Esteves}, {De Mooij}, \&
  {Jayawardhana}}]{2013ApJ...772...51E}
{Esteves}, L.~J., {De Mooij}, E. J.~W., \& {Jayawardhana}, R. 2013, \apj, 772,
  51

\bibitem[{{Foreman-Mackey} {et~al.}(2017){Foreman-Mackey}, {Agol},
  {Ambikasaran}, \& {Angus}}]{2017AJ....154..220F}
{Foreman-Mackey}, D., {Agol}, E., {Ambikasaran}, S., \& {Angus}, R. 2017, \aj,
  154, 220

\bibitem[{{Foreman-Mackey} {et~al.}(2013){Foreman-Mackey}, {Hogg}, {Lang}, \&
  {Goodman}}]{2013PASP..125..306F}
{Foreman-Mackey}, D., {Hogg}, D.~W., {Lang}, D., \& {Goodman}, J. 2013, \pasp,
  125, 306

\bibitem[{{Fulton} \& {Petigura}(2018)}]{2018AJ....156..264F}
{Fulton}, B.~J. \& {Petigura}, E.~A. 2018, \aj, 156, 264

\bibitem[{{Fulton} {et~al.}(2017){Fulton}, {Petigura}, {Howard}, {Isaacson},
  {Marcy}, {Cargile}, {Hebb}, {Weiss}, {Johnson}, {Morton}, {Sinukoff},
  {Crossfield}, \& {Hirsch}}]{2017AJ....154..109F}
{Fulton}, B.~J., {Petigura}, E.~A., {Howard}, A.~W., {et~al.} 2017, \aj, 154,
  109

\bibitem[{{Gaia Collaboration} {et~al.}(2018){Gaia Collaboration}, {Brown},
  {Vallenari}, {Prusti}, {de Bruijne}, {Babusiaux}, {Bailer-Jones}, {Biermann},
  {Evans}, {Eyer}, {Jansen}, {Jordi}, {Klioner}, {Lammers}, {Lindegren},
  {Luri}, {Mignard}, {Panem}, {Pourbaix}, {Randich}, {Sartoretti}, {Siddiqui},
  {Soubiran}, {van Leeuwen}, {Walton}, {Arenou}, {Bastian}, {Cropper},
  {Drimmel}, {Katz}, {Lattanzi}, {Bakker}, {Cacciari}, {Casta{\~n}eda},
  {Chaoul}, {Cheek}, {De Angeli}, {Fabricius}, {Guerra}, {Holl}, {Masana},
  {Messineo}, {Mowlavi}, {Nienartowicz}, {Panuzzo}, {Portell}, {Riello},
  {Seabroke}, {Tanga}, {Th{\'e}venin}, {Gracia-Abril}, {Comoretto},
  {Garcia-Reinaldos}, {Teyssier}, {Altmann}, {Andrae}, {Audard},
  {Bellas-Velidis}, {Benson}, {Berthier}, {Blomme}, {Burgess}, {Busso},
  {Carry}, {Cellino}, {Clementini}, {Clotet}, {Creevey}, {Davidson}, {De
  Ridder}, {Delchambre}, {Dell'Oro}, {Ducourant},
  {Fern{\'a}ndez-Hern{\'a}ndez}, {Fouesneau}, {Fr{\'e}mat}, {Galluccio},
  {Garc{\'\i}a-Torres}, {Gonz{\'a}lez-N{\'u}{\~n}ez}, {Gonz{\'a}lez-Vidal},
  {Gosset}, {Guy}, {Halbwachs}, {Hambly}, {Harrison}, {Hern{\'a}ndez},
  {Hestroffer}, {Hodgkin}, {Hutton}, {Jasniewicz}, {Jean-Antoine-Piccolo},
  {Jordan}, {Korn}, {Krone-Martins}, {Lanzafame}, {Lebzelter}, {L{\"o}ffler},
  {Manteiga}, {Marrese}, {Mart{\'\i}n-Fleitas}, {Moitinho}, {Mora}, {Muinonen},
  {Osinde}, {Pancino}, {Pauwels}, {Petit}, {Recio-Blanco}, {Richards},
  {Rimoldini}, {Robin}, {Sarro}, {Siopis}, {Smith}, {Sozzetti}, {S{\"u}veges},
  {Torra}, {van Reeven}, {Abbas}, {Abreu Aramburu}, {Accart}, {Aerts},
  {Altavilla}, {{\'A}lvarez}, {Alvarez}, {Alves}, {Anderson}, {Andrei},
  {Anglada Varela}, {Antiche}, {Antoja}, {Arcay}, {Astraatmadja}, {Bach},
  {Baker}, {Balaguer-N{\'u}{\~n}ez}, {Balm}, {Barache}, {Barata}, {Barbato},
  {Barblan}, {Barklem}, {Barrado}, {Barros}, {Barstow}, {Bartholom{\'e}
  Mu{\~n}oz}, {Bassilana}, {Becciani}, {Bellazzini}, {Berihuete}, {Bertone},
  {Bianchi}, {Bienaym{\'e}}, {Blanco-Cuaresma}, {Boch}, {Boeche}, {Bombrun},
  {Borrachero}, {Bossini}, {Bouquillon}, {Bourda}, {Bragaglia}, {Bramante},
  {Breddels}, {Bressan}, {Brouillet}, {Br{\"u}semeister}, {Brugaletta},
  {Bucciarelli}, {Burlacu}, {Busonero}, {Butkevich}, {Buzzi}, {Caffau},
  {Cancelliere}, {Cannizzaro}, {Cantat-Gaudin}, {Carballo}, {Carlucci},
  {Carrasco}, {Casamiquela}, {Castellani}, {Castro-Ginard}, {Charlot},
  {Chemin}, {Chiavassa}, {Cocozza}, {Costigan}, {Cowell}, {Crifo}, {Crosta},
  {Crowley}, {Cuypers}, {Dafonte}, {Damerdji}, {Dapergolas}, {David}, {David},
  {de Laverny}, {De Luise}, {De March}, {de Martino}, {de Souza}, {de Torres},
  {Debosscher}, {del Pozo}, {Delbo}, {Delgado}, {Delgado}, {Di Matteo},
  {Diakite}, {Diener}, {Distefano}, {Dolding}, {Drazinos}, {Dur{\'a}n},
  {Edvardsson}, {Enke}, {Eriksson}, {Esquej}, {Eynard Bontemps}, {Fabre},
  {Fabrizio}, {Faigler}, {Falc{\~a}o}, {Farr{\`a}s Casas}, {Federici},
  {Fedorets}, {Fernique}, {Figueras}, {Filippi}, {Findeisen}, {Fonti},
  {Fraile}, {Fraser}, {Fr{\'e}zouls}, {Gai}, {Galleti}, {Garabato},
  {Garc{\'\i}a-Sedano}, {Garofalo}, {Garralda}, {Gavel}, {Gavras}, {Gerssen},
  {Geyer}, {Giacobbe}, {Gilmore}, {Girona}, {Giuffrida}, {Glass}, {Gomes},
  {Granvik}, {Gueguen}, {Guerrier}, {Guiraud}, {Guti{\'e}rrez-S{\'a}nchez},
  {Haigron}, {Hatzidimitriou}, {Hauser}, {Haywood}, {Heiter}, {Helmi}, {Heu},
  {Hilger}, {Hobbs}, {Hofmann}, {Holland}, {Huckle}, {Hypki}, {Icardi},
  {Jan{\ss}en}, {Jevardat de Fombelle}, {Jonker}, {Juh{\'a}sz}, {Julbe},
  {Karampelas}, {Kewley}, {Klar}, {Kochoska}, {Kohley}, {Kolenberg},
  {Kontizas}, {Kontizas}, {Koposov}, {Kordopatis}, {Kostrzewa-Rutkowska},
  {Koubsky}, {Lambert}, {Lanza}, {Lasne}, {Lavigne}, {Le Fustec}, {Le
  Poncin-Lafitte}, {Lebreton}, {Leccia}, {Leclerc}, {Lecoeur-Taibi},
  {Lenhardt}, {Leroux}, {Liao}, {Licata}, {Lindstr{\o}m}, {Lister}, {Livanou},
  {Lobel}, {L{\'o}pez}, {Managau}, {Mann}, {Mantelet}, {Marchal}, {Marchant},
  {Marconi}, {Marinoni}, {Marschalk{\'o}}, {Marshall}, {Martino}, {Marton},
  {Mary}, {Massari}, {Matijevi{\v{c}}}, {Mazeh}, {McMillan}, {Messina},
  {Michalik}, {Millar}, {Molina}, {Molinaro}, {Moln{\'a}r}, {Montegriffo},
  {Mor}, {Morbidelli}, {Morel}, {Morris}, {Mulone}, {Muraveva}, {Musella},
  {Nelemans}, {Nicastro}, {Noval}, {O'Mullane}, {Ord{\'e}novic},
  {Ord{\'o}{\~n}ez-Blanco}, {Osborne}, {Pagani}, {Pagano}, {Pailler},
  {Palacin}, {Palaversa}, {Panahi}, {Pawlak}, {Piersimoni}, {Pineau}, {Plachy},
  {Plum}, {Poggio}, {Poujoulet}, {Pr{\v{s}}a}, {Pulone}, {Racero}, {Ragaini},
  {Rambaux}, {Ramos-Lerate}, {Regibo}, {Reyl{\'e}}, {Riclet}, {Ripepi}, {Riva},
  {Rivard}, {Rixon}, {Roegiers}, {Roelens}, {Romero-G{\'o}mez}, {Rowell},
  {Royer}, {Ruiz-Dern}, {Sadowski}, {Sagrist{\`a} Sell{\'e}s}, {Sahlmann},
  {Salgado}, {Salguero}, {Sanna}, {Santana-Ros}, {Sarasso}, {Savietto},
  {Schultheis}, {Sciacca}, {Segol}, {Segovia}, {S{\'e}gransan}, {Shih},
  {Siltala}, {Silva}, {Smart}, {Smith}, {Solano}, {Solitro}, {Sordo}, {Soria
  Nieto}, {Souchay}, {Spagna}, {Spoto}, {Stampa}, {Steele},
  {Steidelm{\"u}ller}, {Stephenson}, {Stoev}, {Suess}, {Surdej}, {Szabados},
  {Szegedi-Elek}, {Tapiador}, {Taris}, {Tauran}, {Taylor}, {Teixeira},
  {Terrett}, {Teyssand ier}, {Thuillot}, {Titarenko}, {Torra Clotet}, {Turon},
  {Ulla}, {Utrilla}, {Uzzi}, {Vaillant}, {Valentini}, {Valette}, {van Elteren},
  {Van Hemelryck}, {van Leeuwen}, {Vaschetto}, {Vecchiato}, {Veljanoski},
  {Viala}, {Vicente}, {Vogt}, {von Essen}, {Voss}, {Votruba}, {Voutsinas},
  {Walmsley}, {Weiler}, {Wertz}, {Wevers}, {Wyrzykowski}, {Yoldas},
  {{\v{Z}}erjal}, {Ziaeepour}, {Zorec}, {Zschocke}, {Zucker}, {Zurbach}, \&
  {Zwitter}}]{2018A&A...616A...1G}
{Gaia Collaboration}, {Brown}, A.~G.~A., {Vallenari}, A., {et~al.} 2018, \aap,
  616, A1

\bibitem[{{Gaia Collaboration} {et~al.}(2016){Gaia Collaboration}, {Prusti},
  {de Bruijne}, {Brown}, {Vallenari}, {Babusiaux}, {Bailer-Jones}, {Bastian},
  {Biermann}, {Evans}, {Eyer}, {Jansen}, {Jordi}, {Klioner}, {Lammers},
  {Lindegren}, {Luri}, {Mignard}, {Milligan}, {Panem}, {Poinsignon},
  {Pourbaix}, {Randich}, {Sarri}, {Sartoretti}, {Siddiqui}, {Soubiran},
  {Valette}, {van Leeuwen}, {Walton}, {Aerts}, {Arenou}, {Cropper}, {Drimmel},
  {H{\o}g}, {Katz}, {Lattanzi}, {O'Mullane}, {Grebel}, {Holland}, {Huc},
  {Passot}, {Bramante}, {Cacciari}, {Casta{\~n}eda}, {Chaoul}, {Cheek}, {De
  Angeli}, {Fabricius}, {Guerra}, {Hern{\'a}ndez}, {Jean-Antoine-Piccolo},
  {Masana}, {Messineo}, {Mowlavi}, {Nienartowicz}, {Ord{\'o}{\~n}ez-Blanco},
  {Panuzzo}, {Portell}, {Richards}, {Riello}, {Seabroke}, {Tanga},
  {Th{\'e}venin}, {Torra}, {Els}, {Gracia-Abril}, {Comoretto},
  {Garcia-Reinaldos}, {Lock}, {Mercier}, {Altmann}, {Andrae}, {Astraatmadja},
  {Bellas-Velidis}, {Benson}, {Berthier}, {Blomme}, {Busso}, {Carry},
  {Cellino}, {Clementini}, {Cowell}, {Creevey}, {Cuypers}, {Davidson}, {De
  Ridder}, {de Torres}, {Delchambre}, {Dell'Oro}, {Ducourant}, {Fr{\'e}mat},
  {Garc{\'\i}a-Torres}, {Gosset}, {Halbwachs}, {Hambly}, {Harrison}, {Hauser},
  {Hestroffer}, {Hodgkin}, {Huckle}, {Hutton}, {Jasniewicz}, {Jordan},
  {Kontizas}, {Korn}, {Lanzafame}, {Manteiga}, {Moitinho}, {Muinonen},
  {Osinde}, {Pancino}, {Pauwels}, {Petit}, {Recio-Blanco}, {Robin}, {Sarro},
  {Siopis}, {Smith}, {Smith}, {Sozzetti}, {Thuillot}, {van Reeven}, {Viala},
  {Abbas}, {Abreu Aramburu}, {Accart}, {Aguado}, {Allan}, {Allasia},
  {Altavilla}, {{\'A}lvarez}, {Alves}, {Anderson}, {Andrei}, {Anglada Varela},
  {Antiche}, {Antoja}, {Ant{\'o}n}, {Arcay}, {Atzei}, {Ayache}, {Bach},
  {Baker}, {Balaguer-N{\'u}{\~n}ez}, {Barache}, {Barata}, {Barbier}, {Barblan},
  {Baroni}, {Barrado y Navascu{\'e}s}, {Barros}, {Barstow}, {Becciani},
  {Bellazzini}, {Bellei}, {Bello Garc{\'\i}a}, {Belokurov}, {Bendjoya},
  {Berihuete}, {Bianchi}, {Bienaym{\'e}}, {Billebaud}, {Blagorodnova},
  {Blanco-Cuaresma}, {Boch}, {Bombrun}, {Borrachero}, {Bouquillon}, {Bourda},
  {Bouy}, {Bragaglia}, {Breddels}, {Brouillet}, {Br{\"u}semeister},
  {Bucciarelli}, {Budnik}, {Burgess}, {Burgon}, {Burlacu}, {Busonero}, {Buzzi},
  {Caffau}, {Cambras}, {Campbell}, {Cancelliere}, {Cantat-Gaudin}, {Carlucci},
  {Carrasco}, {Castellani}, {Charlot}, {Charnas}, {Charvet}, {Chassat},
  {Chiavassa}, {Clotet}, {Cocozza}, {Collins}, {Collins}, {Costigan}, {Crifo},
  {Cross}, {Crosta}, {Crowley}, {Dafonte}, {Damerdji}, {Dapergolas}, {David},
  {David}, {De Cat}, {de Felice}, {de Laverny}, {De Luise}, {De March}, {de
  Martino}, {de Souza}, {Debosscher}, {del Pozo}, {Delbo}, {Delgado},
  {Delgado}, {di Marco}, {Di Matteo}, {Diakite}, {Distefano}, {Dolding}, {Dos
  Anjos}, {Drazinos}, {Dur{\'a}n}, {Dzigan}, {Ecale}, {Edvardsson}, {Enke},
  {Erdmann}, {Escolar}, {Espina}, {Evans}, {Eynard Bontemps}, {Fabre},
  {Fabrizio}, {Faigler}, {Falc{\~a}o}, {Farr{\`a}s Casas}, {Faye}, {Federici},
  {Fedorets}, {Fern{\'a}ndez-Hern{\'a}ndez}, {Fernique}, {Fienga}, {Figueras},
  {Filippi}, {Findeisen}, {Fonti}, {Fouesneau}, {Fraile}, {Fraser}, {Fuchs},
  {Furnell}, {Gai}, {Galleti}, {Galluccio}, {Garabato}, {Garc{\'\i}a-Sedano},
  {Gar{\'e}}, {Garofalo}, {Garralda}, {Gavras}, {Gerssen}, {Geyer}, {Gilmore},
  {Girona}, {Giuffrida}, {Gomes}, {Gonz{\'a}lez-Marcos},
  {Gonz{\'a}lez-N{\'u}{\~n}ez}, {Gonz{\'a}lez-Vidal}, {Granvik}, {Guerrier},
  {Guillout}, {Guiraud}, {G{\'u}rpide}, {Guti{\'e}rrez-S{\'a}nchez}, {Guy},
  {Haigron}, {Hatzidimitriou}, {Haywood}, {Heiter}, {Helmi}, {Hobbs},
  {Hofmann}, {Holl}, {Holland }, {Hunt}, {Hypki}, {Icardi}, {Irwin}, {Jevardat
  de Fombelle}, {Jofr{\'e}}, {Jonker}, {Jorissen}, {Julbe}, {Karampelas},
  {Kochoska}, {Kohley}, {Kolenberg}, {Kontizas}, {Koposov}, {Kordopatis},
  {Koubsky}, {Kowalczyk}, {Krone-Martins}, {Kudryashova}, {Kull}, {Bachchan},
  {Lacoste-Seris}, {Lanza}, {Lavigne}, {Le Poncin-Lafitte}, {Lebreton},
  {Lebzelter}, {Leccia}, {Leclerc}, {Lecoeur-Taibi}, {Lemaitre}, {Lenhardt},
  {Leroux}, {Liao}, {Licata}, {Lindstr{\o}m}, {Lister}, {Livanou}, {Lobel},
  {L{\"o}ffler}, {L{\'o}pez}, {Lopez-Lozano}, {Lorenz}, {Loureiro},
  {MacDonald}, {Magalh{\~a}es Fernandes}, {Managau}, {Mann}, {Mantelet},
  {Marchal}, {Marchant}, {Marconi}, {Marie}, {Marinoni}, {Marrese},
  {Marschalk{\'o}}, {Marshall}, {Mart{\'\i}n-Fleitas}, {Martino}, {Mary},
  {Matijevi{\v{c}}}, {Mazeh}, {McMillan}, {Messina}, {Mestre}, {Michalik},
  {Millar}, {Miranda}, {Molina}, {Molinaro}, {Molinaro}, {Moln{\'a}r},
  {Moniez}, {Montegriffo}, {Monteiro}, {Mor}, {Mora}, {Morbidelli}, {Morel},
  {Morgenthaler}, {Morley}, {Morris}, {Mulone}, {Muraveva}, {Musella},
  {Narbonne}, {Nelemans}, {Nicastro}, {Noval}, {Ord{\'e}novic},
  {Ordieres-Mer{\'e}}, {Osborne}, {Pagani}, {Pagano}, {Pailler}, {Palacin},
  {Palaversa}, {Parsons}, {Paulsen}, {Pecoraro}, {Pedrosa}, {Pentik{\"a}inen},
  {Pereira}, {Pichon}, {Piersimoni}, {Pineau}, {Plachy}, {Plum}, {Poujoulet},
  {Pr{\v{s}}a}, {Pulone}, {Ragaini}, {Rago}, {Rambaux}, {Ramos-Lerate},
  {Ranalli}, {Rauw}, {Read}, {Regibo}, {Renk}, {Reyl{\'e}}, {Ribeiro},
  {Rimoldini}, {Ripepi}, {Riva}, {Rixon}, {Roelens}, {Romero-G{\'o}mez},
  {Rowell}, {Royer}, {Rudolph}, {Ruiz-Dern}, {Sadowski}, {Sagrist{\`a}
  Sell{\'e}s}, {Sahlmann}, {Salgado}, {Salguero}, {Sarasso}, {Savietto},
  {Schnorhk}, {Schultheis}, {Sciacca}, {Segol}, {Segovia}, {Segransan},
  {Serpell}, {Shih}, {Smareglia}, {Smart}, {Smith}, {Solano}, {Solitro},
  {Sordo}, {Soria Nieto}, {Souchay}, {Spagna}, {Spoto}, {Stampa}, {Steele},
  {Steidelm{\"u}ller}, {Stephenson}, {Stoev}, {Suess}, {S{\"u}veges}, {Surdej},
  {Szabados}, {Szegedi-Elek}, {Tapiador}, {Taris}, {Tauran}, {Taylor},
  {Teixeira}, {Terrett}, {Tingley}, {Trager}, {Turon}, {Ulla}, {Utrilla},
  {Valentini}, {van Elteren}, {Van Hemelryck}, {van Leeuwen}, {Varadi},
  {Vecchiato}, {Veljanoski}, {Via}, {Vicente}, {Vogt}, {Voss}, {Votruba},
  {Voutsinas}, {Walmsley}, {Weiler}, {Weingrill}, {Werner}, {Wevers},
  {Whitehead}, {Wyrzykowski}, {Yoldas}, {{\v{Z}}erjal}, {Zucker}, {Zurbach},
  {Zwitter}, {Alecu}, {Allen}, {Allende Prieto}, {Amorim},
  {Anglada-Escud{\'e}}, {Arsenijevic}, {Azaz}, {Balm}, {Beck}, {Bernstein},
  {Bigot}, {Bijaoui}, {Blasco}, {Bonfigli}, {Bono}, {Boudreault}, {Bressan},
  {Brown}, {Brunet}, {Bunclark}, {Buonanno}, {Butkevich}, {Carret}, {Carrion},
  {Chemin}, {Ch{\'e}reau}, {Corcione}, {Darmigny}, {de Boer}, {de Teodoro}, {de
  Zeeuw}, {Delle Luche}, {Domingues}, {Dubath}, {Fodor}, {Fr{\'e}zouls},
  {Fries}, {Fustes}, {Fyfe}, {Gallardo}, {Gallegos}, {Gardiol}, {Gebran},
  {Gomboc}, {G{\'o}mez}, {Grux}, {Gueguen}, {Heyrovsky}, {Hoar}, {Iannicola},
  {Isasi Parache}, {Janotto}, {Joliet}, {Jonckheere}, {Keil}, {Kim},
  {Klagyivik}, {Klar}, {Knude}, {Kochukhov}, {Kolka}, {Kos}, {Kutka}, {Lainey},
  {LeBouquin}, {Liu}, {Loreggia}, {Makarov}, {Marseille}, {Martayan},
  {Martinez-Rubi}, {Massart}, {Meynadier}, {Mignot}, {Munari}, {Nguyen},
  {Nordlander}, {Ocvirk}, {O'Flaherty}, {Olias Sanz}, {Ortiz}, {Osorio},
  {Oszkiewicz}, {Ouzounis}, {Palmer}, {Park}, {Pasquato}, {Peltzer}, {Peralta},
  {P{\'e}turaud}, {Pieniluoma}, {Pigozzi}, {Poels}, {Prat}, {Prod'homme},
  {Raison}, {Rebordao}, {Risquez}, {Rocca-Volmerange}, {Rosen}, {Ruiz-Fuertes},
  {Russo}, {Sembay}, {Serraller Vizcaino}, {Short}, {Siebert}, {Silva},
  {Sinachopoulos}, {Slezak}, {Soffel}, {Sosnowska}, {Strai{\v{z}}ys}, {ter
  Linden}, {Terrell}, {Theil}, {Tiede}, {Troisi}, {Tsalmantza}, {Tur},
  {Vaccari}, {Vachier}, {Valles}, {Van Hamme}, {Veltz}, {Virtanen}, {Wallut},
  {Wichmann}, {Wilkinson}, {Ziaeepour}, \& {Zschocke}}]{2016A&A...595A...1G}
{Gaia Collaboration}, {Prusti}, T., {de Bruijne}, J.~H.~J., {et~al.} 2016,
  \aap, 595, A1

\bibitem[{{Grunblatt} {et~al.}(2015){Grunblatt}, {Howard}, \&
  {Haywood}}]{2015ApJ...808..127G}
{Grunblatt}, S.~K., {Howard}, A.~W., \& {Haywood}, R.~D. 2015, \apj, 808, 127

\bibitem[{{Howard} {et~al.}(2012){Howard}, {Marcy}, {Bryson}, {Jenkins},
  {Rowe}, {Batalha}, {Borucki}, {Koch}, {Dunham}, {Gautier}, {Van Cleve},
  {Cochran}, {Latham}, {Lissauer}, {Torres}, {Brown}, {Gilliland}, {Buchhave},
  {Caldwell}, {Christensen-Dalsgaard}, {Ciardi}, {Fressin}, {Haas}, {Howell},
  {Kjeldsen}, {Seager}, {Rogers}, {Sasselov}, {Steffen}, {Basri},
  {Charbonneau}, {Christiansen}, {Clarke}, {Dupree}, {Fabrycky}, {Fischer},
  {Ford}, {Fortney}, {Tarter}, {Girouard}, {Holman}, {Johnson}, {Klaus},
  {Machalek}, {Moorhead}, {Morehead}, {Ragozzine}, {Tenenbaum}, {Twicken},
  {Quinn}, {Isaacson}, {Shporer}, {Lucas}, {Walkowicz}, {Welsh}, {Boss},
  {Devore}, {Gould}, {Smith}, {Morris}, {Prsa}, {Morton}, {Still}, {Thompson},
  {Mullally}, {Endl}, \& {MacQueen}}]{2012ApJS..201...15H}
{Howard}, A.~W., {Marcy}, G.~W., {Bryson}, S.~T., {et~al.} 2012, \apjs, 201, 15

\bibitem[{Kass \& Raftery(1995)}]{KassRaftery}
Kass, R.~E. \& Raftery, A.~E. 1995, Journal of the American Statistical
  Association, 90, 773

\bibitem[{{Kipping}(2010)}]{2010MNRAS.408.1758K}
{Kipping}, D.~M. 2010, \mnras, 408, 1758

\bibitem[{{Kipping}(2013)}]{2013MNRAS.435.2152K}
{Kipping}, D.~M. 2013, \mnras, 435, 2152

\bibitem[{{Kite} {et~al.}(2016){Kite}, {Fegley}, {Schaefer}, \&
  {Gaidos}}]{2016ApJ...828...80K}
{Kite}, E.~S., {Fegley}, Jr., B., {Schaefer}, L., \& {Gaidos}, E. 2016, \apj,
  828, 80

\bibitem[{{Kov{\'a}cs} {et~al.}(2002){Kov{\'a}cs}, {Zucker}, \&
  {Mazeh}}]{2002A&A...391..369K}
{Kov{\'a}cs}, G., {Zucker}, S., \& {Mazeh}, T. 2002, \aap, 391, 369

\bibitem[{{Kreidberg}(2015)}]{2015PASP..127.1161K}
{Kreidberg}, L. 2015, \pasp, 127, 1161

\bibitem[{Kreidberg {et~al.}(2019)Kreidberg, Koll, Morley, Hu, Schaefer,
  Deming, Stevenson, Dittmann, Vanderburg, Berardo, Guo, Stassun, Crossfield,
  Charbonneau, Latham, Loeb, Ricker, Seager, \& Vanderspek}]{Kreidberg2019}
Kreidberg, L., Koll, D. D.~B., Morley, C., {et~al.} 2019, Nature, 573, 87

\bibitem[{{Kurucz}(1993)}]{1993sssp.book.....K}
{Kurucz}, R.~L. 1993, {SYNTHE spectrum synthesis programs and line data}

\bibitem[{{Lee} \& {Chiang}(2017)}]{2017ApJ...842...40L}
{Lee}, E.~J. \& {Chiang}, E. 2017, \apj, 842, 40

\bibitem[{{L{\'e}ger} {et~al.}(2011){L{\'e}ger}, {Grasset}, {Fegley}, {Codron},
  {Albarede}, {Barge}, {Barnes}, {Cance}, {Carpy}, {Catalano}, {Cavarroc},
  {Demangeon}, {Ferraz-Mello}, {Gabor}, {Grie{\ss}meier}, {Leibacher},
  {Libourel}, {Maurin}, {Raymond}, {Rouan}, {Samuel}, {Schaefer}, {Schneider},
  {Schuller}, {Selsis}, \& {Sotin}}]{2011Icar..213....1L}
{L{\'e}ger}, A., {Grasset}, O., {Fegley}, B., {et~al.} 2011, \icarus, 213, 1

\bibitem[{Lewis(2004)}]{Lewis2004}
Lewis, J.~S. 2004, {Physics and chemistry of the solar system} (Elsevier
  Academic Press), 655

\bibitem[{{Liddle}(2007)}]{2007MNRAS.377L..74L}
{Liddle}, A.~R. 2007, \mnras, 377, L74

\bibitem[{{Lopez} \& {Fortney}(2013)}]{2013ApJ...776....2L}
{Lopez}, E.~D. \& {Fortney}, J.~J. 2013, \apj, 776, 2

\bibitem[{{L{\'o}pez-Morales} {et~al.}(2016){L{\'o}pez-Morales}, {Haywood},
  {Coughlin}, {Zeng}, {Buchhave}, {Giles}, {Affer}, {Bonomo}, {Charbonneau},
  {Collier Cameron}, {Consentino}, {Dressing}, {Dumusque}, {Figueira},
  {Fiorenzano}, {Harutyunyan}, {Johnson}, {Latham}, {Lopez}, {Lovis},
  {Malavolta}, {Mayor}, {Micela}, {Molinari}, {Mortier}, {Motalebi},
  {Nascimbeni}, {Pepe}, {Phillips}, {Piotto}, {Pollacco}, {Queloz}, {Rice},
  {Sasselov}, {Segransan}, {Sozzetti}, {Udry}, {Vanderburg}, \&
  {Watson}}]{2016AJ....152..204L}
{L{\'o}pez-Morales}, M., {Haywood}, R.~D., {Coughlin}, J.~L., {et~al.} 2016,
  \aj, 152, 204

\bibitem[{{Lundkvist} {et~al.}(2016){Lundkvist}, {Kjeldsen}, {Albrecht},
  {Davies}, {Basu}, {Huber}, {Justesen}, {Karoff}, {Silva Aguirre}, {van
  Eylen}, {Vang}, {Arentoft}, {Barclay}, {Bedding}, {Campante}, {Chaplin},
  {Christensen-Dalsgaard}, {Elsworth}, {Gilliland}, {Handberg}, {Hekker},
  {Kawaler}, {Lund}, {Metcalfe}, {Miglio}, {Rowe}, {Stello}, {Tingley}, \&
  {White}}]{2016NatCo...711201L}
{Lundkvist}, M.~S., {Kjeldsen}, H., {Albrecht}, S., {et~al.} 2016, Nature
  Communications, 7, 11201

\bibitem[{{Malavolta} {et~al.}(2017){Malavolta}, {Lovis}, {Pepe}, {Sneden}, \&
  {Udry}}]{2017MNRAS.469.3965M}
{Malavolta}, L., {Lovis}, C., {Pepe}, F., {Sneden}, C., \& {Udry}, S. 2017,
  \mnras, 469, 3965

\bibitem[{{Malavolta} {et~al.}(2018){Malavolta}, {Mayo}, {Louden}, {Rajpaul},
  {Bonomo}, {Buchhave}, {Kreidberg}, {Kristiansen}, {Lopez-Morales}, {Mortier},
  {Vand erburg}, {Coffinet}, {Ehrenreich}, {Lovis}, {Bouchy}, {Charbonneau},
  {Ciardi}, {Collier Cameron}, {Cosentino}, {Crossfield}, {Damasso},
  {Dressing}, {Dumusque}, {Everett}, {Figueira}, {Fiorenzano}, {Gonzales},
  {Haywood}, {Harutyunyan}, {Hirsch}, {Howell}, {Johnson}, {Latham}, {Lopez},
  {Mayor}, {Micela}, {Molinari}, {Nascimbeni}, {Pepe}, {Phillips}, {Piotto},
  {Rice}, {Sasselov}, {S{\'e}gransan}, {Sozzetti}, {Udry}, \&
  {Watson}}]{2018AJ....155..107M}
{Malavolta}, L., {Mayo}, A.~W., {Louden}, T., {et~al.} 2018, \aj, 155, 107

\bibitem[{{Malavolta} {et~al.}(2016){Malavolta}, {Nascimbeni}, {Piotto},
  {Quinn}, {Borsato}, {Granata}, {Bonomo}, {Marzari}, {Bedin}, {Rainer},
  {Desidera}, {Lanza}, {Poretti}, {Sozzetti}, {White}, {Latham}, {Cunial},
  {Libralato}, {Nardiello}, {Boccato}, {Claudi}, {Cosentino}, {Covino},
  {Gratton}, {Maggio}, {Micela}, {Molinari}, {Pagano}, {Smareglia}, {Affer},
  {Andreuzzi}, {Aparicio}, {Benatti}, {Bignamini}, {Borsa}, {Damasso}, {Di
  Fabrizio}, {Harutyunyan}, {Esposito}, {Fiorenzano}, {Gandolfi}, {Giacobbe},
  {Gonz{\'a}lez Hern{\'a}ndez}, {Maldonado}, {Masiero}, {Molinaro}, {Pedani},
  \& {Scandariato}}]{2016A&A...588A.118M}
{Malavolta}, L., {Nascimbeni}, V., {Piotto}, G., {et~al.} 2016, \aap, 588, A118

\bibitem[{{Mandel} \& {Agol}(2002)}]{2002ApJ...580L.171M}
{Mandel}, K. \& {Agol}, E. 2002, \apj, 580, L171

\bibitem[{{Marcus} {et~al.}(2010){Marcus}, {Sasselov}, {Hernquist}, \&
  {Stewart}}]{2010ApJ...712L..73M}
{Marcus}, R.~A., {Sasselov}, D., {Hernquist}, L., \& {Stewart}, S.~T. 2010,
  \apjl, 712, L73

\bibitem[{{Matsumura} {et~al.}(2008){Matsumura}, {Takeda}, \&
  {Rasio}}]{2008ApJ...686L..29M}
{Matsumura}, S., {Takeda}, G., \& {Rasio}, F.~A. 2008, \apjl, 686, L29

\bibitem[{{Mayor} {et~al.}(2011){Mayor}, {Marmier}, {Lovis}, {Udry},
  {S{\'e}gransan}, {Pepe}, {Benz}, {Bertaux}, {Bouchy}, {Dumusque}, {Lo Curto},
  {Mordasini}, {Queloz}, \& {Santos}}]{2011arXiv1109.2497M}
{Mayor}, M., {Marmier}, M., {Lovis}, C., {et~al.} 2011, arXiv e-prints,
  arXiv:1109.2497

\bibitem[{{Mortier} {et~al.}(2014){Mortier}, {Sousa}, {Adibekyan}, {Brand
  {\~a}o}, \& {Santos}}]{2014A&A...572A..95M}
{Mortier}, A., {Sousa}, S.~G., {Adibekyan}, V.~Z., {Brand {\~a}o}, I.~M., \&
  {Santos}, N.~C. 2014, \aap, 572, A95

\bibitem[{{Morton}(2015)}]{2015ascl.soft03010M}
{Morton}, T.~D. 2015, {isochrones: Stellar model grid package}, Astrophysics
  Source Code Library, record 1503.010

\bibitem[{{Owen} \& {Wu}(2013)}]{2013ApJ...775..105O}
{Owen}, J.~E. \& {Wu}, Y. 2013, \apj, 775, 105

\bibitem[{{Owen} \& {Wu}(2017)}]{2017ApJ...847...29O}
{Owen}, J.~E. \& {Wu}, Y. 2017, \apj, 847, 29

\bibitem[{{Paxton} {et~al.}(2011){Paxton}, {Bildsten}, {Dotter}, {Herwig},
  {Lesaffre}, \& {Timmes}}]{2011ApJS..192....3P}
{Paxton}, B., {Bildsten}, L., {Dotter}, A., {et~al.} 2011, \apjs, 192, 3

\bibitem[{{Pepe} {et~al.}(2013){Pepe}, {Cameron}, {Latham}, {Molinari}, {Udry},
  {Bonomo}, {Buchhave}, {Charbonneau}, {Cosentino}, {Dressing}, {Dumusque},
  {Figueira}, {Fiorenzano}, {Gettel}, {Harutyunyan}, {Haywood}, {Horne},
  {Lopez-Morales}, {Lovis}, {Malavolta}, {Mayor}, {Micela}, {Motalebi},
  {Nascimbeni}, {Phillips}, {Piotto}, {Pollacco}, {Queloz}, {Rice}, {Sasselov},
  {S{\'e}gransan}, {Sozzetti}, {Szentgyorgyi}, \&
  {Watson}}]{2013Natur.503..377P}
{Pepe}, F., {Cameron}, A.~C., {Latham}, D.~W., {et~al.} 2013, \nat, 503, 377

\bibitem[{{Pepe} {et~al.}(2002){Pepe}, {Mayor}, {Rupprecht}, {Avila},
  {Ballester}, {Beckers}, {Benz}, {Bertaux}, {Bouchy}, {Buzzoni}, {Cavadore},
  {Deiries}, {Dekker}, {Delabre}, {D'Odorico}, {Eckert}, {Fischer}, {Fleury},
  {George}, {Gilliotte}, {Gojak}, {Guzman}, {Koch}, {Kohler}, {Kotzlowski},
  {Lacroix}, {Le Merrer}, {Lizon}, {Lo Curto}, {Longinotti}, {Megevand},
  {Pasquini}, {Petitpas}, {Pichard}, {Queloz}, {Reyes}, {Richaud}, {Sivan},
  {Sosnowska}, {Soto}, {Udry}, {Ureta}, {van Kesteren}, {Weber}, {Weilenmann},
  {Wicenec}, {Wieland}, {Christensen-Dalsgaard}, {Dravins}, {Hatzes},
  {K{\"u}rster}, {Paresce}, \& {Penny}}]{2002Msngr.110....9P}
{Pepe}, F., {Mayor}, M., {Rupprecht}, G., {et~al.} 2002, The Messenger, 110, 9

\bibitem[{{Pont} {et~al.}(2006){Pont}, {Zucker}, \&
  {Queloz}}]{2006MNRAS.373..231P}
{Pont}, F., {Zucker}, S., \& {Queloz}, D. 2006, \mnras, 373, 231

\bibitem[{{Queloz} {et~al.}(2009){Queloz}, {Bouchy}, {Moutou}, {Hatzes},
  {H{\'e}brard}, {Alonso}, {Auvergne}, {Baglin}, {Barbieri}, {Barge}, {Benz},
  {Bord{\'e}}, {Deeg}, {Deleuil}, {Dvorak}, {Erikson}, {Ferraz Mello},
  {Fridlund}, {Gandolfi}, {Gillon}, {Guenther}, {Guillot}, {Jorda}, {Hartmann},
  {Lammer}, {L{\'e}ger}, {Llebaria}, {Lovis}, {Magain}, {Mayor}, {Mazeh},
  {Ollivier}, {P{\"a}tzold}, {Pepe}, {Rauer}, {Rouan}, {Schneider},
  {Segransan}, {Udry}, \& {Wuchterl}}]{2009A&A...506..303Q}
{Queloz}, D., {Bouchy}, F., {Moutou}, C., {et~al.} 2009, \aap, 506, 303

\bibitem[{{Rice}(2015)}]{2015MNRAS.448.1729R}
{Rice}, K. 2015, \mnras, 448, 1729

\bibitem[{{Rice} {et~al.}(2019){Rice}, {Malavolta}, {Mayo}, {Mortier},
  {Buchhave}, {Affer}, {Vanderburg}, {Lopez-Morales}, {Poretti}, {Zeng},
  {Cameron}, {Damasso}, {Coffinet}, {Latham}, {Bonomo}, {Bouchy},
  {Charbonneau}, {Dumusque}, {Figueira}, {Martinez Fiorenzano}, {Haywood},
  {Johnson}, {Lopez}, {Lovis}, {Mayor}, {Micela}, {Molinari}, {Nascimbeni},
  {Nava}, {Pepe}, {Phillips}, {Piotto}, {Sasselov}, {S{\'e}gransan},
  {Sozzetti}, {Udry}, \& {Watson}}]{2019MNRAS.484.3731R}
{Rice}, K., {Malavolta}, L., {Mayo}, A., {et~al.} 2019, \mnras, 484, 3731

\bibitem[{{Rogers}(2015)}]{2015ApJ...801...41R}
{Rogers}, L.~A. 2015, \apj, 801, 41

\bibitem[{{Sanchis-Ojeda} {et~al.}(2014){Sanchis-Ojeda}, {Rappaport}, {Winn},
  {Kotson}, {Levine}, \& {El Mellah}}]{2014ApJ...787...47S}
{Sanchis-Ojeda}, R., {Rappaport}, S., {Winn}, J.~N., {et~al.} 2014, \apj, 787,
  47

\bibitem[{{Sanchis-Ojeda} {et~al.}(2013){Sanchis-Ojeda}, {Rappaport}, {Winn},
  {Levine}, {Kotson}, {Latham}, \& {Buchhave}}]{2013ApJ...774...54S}
{Sanchis-Ojeda}, R., {Rappaport}, S., {Winn}, J.~N., {et~al.} 2013, \apj, 774,
  54

\bibitem[{{Santerne} {et~al.}(2018){Santerne}, {Brugger}, {Armstrong},
  {Adibekyan}, {Lillo-Box}, {Gosselin}, {Aguichine}, {Almenara}, {Barrado},
  {Barros}, {Bayliss}, {Boisse}, {Bonomo}, {Bouchy}, {Brown}, {Deleuil},
  {Delgado Mena}, {Demangeon}, {D{\'\i}az}, {Doyle}, {Dumusque}, {Faedi},
  {Faria}, {Figueira}, {Foxell}, {Giles}, {H{\'e}brard}, {Hojjatpanah},
  {Hobson}, {Jackman}, {King}, {Kirk}, {Lam}, {Ligi}, {Lovis}, {Louden},
  {McCormac}, {Mousis}, {Neal}, {Osborn}, {Pepe}, {Pollacco}, {Santos},
  {Sousa}, {Udry}, \& {Vigan}}]{2018NatAs...2..393S}
{Santerne}, A., {Brugger}, B., {Armstrong}, D.~J., {et~al.} 2018, Nature
  Astronomy, 2, 393

\bibitem[{Schaefer \& Fegley(2009)}]{Schaefer_2009_ApJ_703_L113}
Schaefer, L. \& Fegley, B. 2009, The Astrophysical Journal, 703, L113

\bibitem[{{Schneider} {et~al.}(2011){Schneider}, {Dedieu}, {Le Sidaner},
  {Savalle}, \& {Zolotukhin}}]{2011A&A...532A..79S}
{Schneider}, J., {Dedieu}, C., {Le Sidaner}, P., {Savalle}, R., \&
  {Zolotukhin}, I. 2011, \aap, 532, A79

\bibitem[{{Skrutskie} {et~al.}(2006){Skrutskie}, {Cutri}, {Stiening},
  {Weinberg}, {Schneider}, {Carpenter}, {Beichman}, {Capps}, {Chester},
  {Elias}, {Huchra}, {Liebert}, {Lonsdale}, {Monet}, {Price}, {Seitzer},
  {Jarrett}, {Kirkpatrick}, {Gizis}, {Howard}, {Evans}, {Fowler}, {Fullmer},
  {Hurt}, {Light}, {Kopan}, {Marsh}, {McCallon}, {Tam}, {Van Dyk}, \&
  {Wheelock}}]{2006AJ....131.1163S}
{Skrutskie}, M.~F., {Cutri}, R.~M., {Stiening}, R., {et~al.} 2006, \aj, 131,
  1163

\bibitem[{{Sneden}(1973)}]{1973PhDT.......180S}
{Sneden}, C.~A. 1973, PhD thesis, The University of Texas at Austin.

\bibitem[{{Sousa}(2014)}]{2014dapb.book..297S}
{Sousa}, S.~G. 2014, {ARES + MOOG: A Practical Overview of an Equivalent Width
  (EW) Method to Derive Stellar Parameters}, 297--310

\bibitem[{{Sousa} {et~al.}(2015){Sousa}, {Santos}, {Adibekyan}, {Delgado-Mena},
  \& {Israelian}}]{2015A&A...577A..67S}
{Sousa}, S.~G., {Santos}, N.~C., {Adibekyan}, V., {Delgado-Mena}, E., \&
  {Israelian}, G. 2015, \aap, 577, A67

\bibitem[{{Sousa} {et~al.}(2011){Sousa}, {Santos}, {Israelian}, {Lovis},
  {Mayor}, {Silva}, \& {Udry}}]{2011A&A...526A..99S}
{Sousa}, S.~G., {Santos}, N.~C., {Israelian}, G., {et~al.} 2011, \aap, 526, A99

\bibitem[{{Sozzetti} \& {de Bruijne}(2018)}]{2018haex.bookE..81S}
{Sozzetti}, A. \& {de Bruijne}, J. 2018, {Space Astrometry Missions for
  Exoplanet Science: Gaia and the Legacy of Hipparcos}, 81

\bibitem[{{Stassun} \& {Torres}(2018)}]{2018ApJ...862...61S}
{Stassun}, K.~G. \& {Torres}, G. 2018, \apj, 862, 61

\bibitem[{{Ter Braak}(2006)}]{2006S&C....16..239T}
{Ter Braak}, C. J.~F. 2006, Statistics and Computing, 16, 239

\bibitem[{{Torres} {et~al.}(2012){Torres}, {Fischer}, {Sozzetti}, {Buchhave},
  {Winn}, {Holman}, \& {Carter}}]{2012ApJ...757..161T}
{Torres}, G., {Fischer}, D.~A., {Sozzetti}, A., {et~al.} 2012, \apj, 757, 161

\bibitem[{{Van Eylen} {et~al.}(2018){Van Eylen}, {Agentoft}, {Lundkvist},
  {Kjeldsen}, {Owen}, {Fulton}, {Petigura}, \& {Snellen}}]{2018MNRAS.479.4786V}
{Van Eylen}, V., {Agentoft}, C., {Lundkvist}, M.~S., {et~al.} 2018, \mnras,
  479, 4786

\bibitem[{{Vanderburg} \& {Johnson}(2014)}]{2014PASP..126..948V}
{Vanderburg}, A. \& {Johnson}, J.~A. 2014, \pasp, 126, 948

\bibitem[{{Vanderburg} {et~al.}(2016){Vanderburg}, {Latham}, {Buchhave},
  {Bieryla}, {Berlind}, {Calkins}, {Esquerdo}, {Welsh}, \&
  {Johnson}}]{2016ApJS..222...14V}
{Vanderburg}, A., {Latham}, D.~W., {Buchhave}, L.~A., {et~al.} 2016, \apjs,
  222, 14

\bibitem[{{Van{\'\i}{\v{c}}ek}(1971)}]{1971Ap&SS..12...10V}
{Van{\'\i}{\v{c}}ek}, P. 1971, \apss, 12, 10

\bibitem[{{Winn} {et~al.}(2009){Winn}, {Johnson}, {Albrecht}, {Howard},
  {Marcy}, {Crossfield}, \& {Holman}}]{2009ApJ...703L..99W}
{Winn}, J.~N., {Johnson}, J.~A., {Albrecht}, S., {et~al.} 2009, \apjl, 703, L99

\bibitem[{{Winn} {et~al.}(2018){Winn}, {Sanchis-Ojeda}, \&
  {Rappaport}}]{2018NewAR..83...37W}
{Winn}, J.~N., {Sanchis-Ojeda}, R., \& {Rappaport}, S. 2018, \nar, 83, 37

\bibitem[{{Winn} {et~al.}(2017){Winn}, {Sanchis-Ojeda}, {Rogers}, {Petigura},
  {Howard}, {Isaacson}, {Marcy}, {Schlaufman}, {Cargile}, \&
  {Hebb}}]{2017AJ....154...60W}
{Winn}, J.~N., {Sanchis-Ojeda}, R., {Rogers}, L., {et~al.} 2017, \aj, 154, 60

\bibitem[{{Wright} {et~al.}(2010){Wright}, {Eisenhardt}, {Mainzer}, {Ressler},
  {Cutri}, {Jarrett}, {Kirkpatrick}, {Padgett}, {McMillan}, {Skrutskie},
  {Stanford}, {Cohen}, {Walker}, {Mather}, {Leisawitz}, {Gautier}, {McLean},
  {Benford}, {Lonsdale}, {Blain}, {Mendez}, {Irace}, {Duval}, {Liu}, {Royer},
  {Heinrichsen}, {Howard}, {Shannon}, {Kendall}, {Walsh}, {Larsen}, {Cardon},
  {Schick}, {Schwalm}, {Abid}, {Fabinsky}, {Naes}, \&
  {Tsai}}]{2010AJ....140.1868W}
{Wright}, E.~L., {Eisenhardt}, P. R.~M., {Mainzer}, A.~K., {et~al.} 2010, \aj,
  140, 1868

\bibitem[{{Yu} {et~al.}(2018){Yu}, {Crossfield}, {Schlieder}, {Kosiarek},
  {Feinstein}, {Livingston}, {Howard}, {Benneke}, {Petigura}, {Bristow},
  {Christiansen}, {Ciardi}, {Crepp}, {Dressing}, {Fulton}, {Gonzales},
  {Hardegree-Ullman}, {Henning}, {Isaacson}, {L{\'e}pine}, {Martinez},
  {Morales}, \& {Sinukoff}}]{2018AJ....156...22Y}
{Yu}, L., {Crossfield}, I. J.~M., {Schlieder}, J.~E., {et~al.} 2018, \aj, 156,
  22

\bibitem[{{Zechmeister} \& {K{\"u}rster}(2009)}]{2009A&A...496..577Z}
{Zechmeister}, M. \& {K{\"u}rster}, M. 2009, \aap, 496, 577

\bibitem[{{Zeng} {et~al.}(2017){Zeng}, {Jacobsen}, \&
  {Sasselov}}]{2017RNAAS...1a..32Z}
{Zeng}, L., {Jacobsen}, S.~B., \& {Sasselov}, D.~D. 2017, Research Notes of the
  American Astronomical Society, 1, 32

\bibitem[{Zeng {et~al.}(2019)Zeng, Jacobsen, Sasselov, Petaev, Vanderburg,
  Lopez-Morales, Perez-Mercader, Mattsson, Li, Heising, Bonomo, Damasso,
  Berger, Cao, Levi, \& Wordsworth}]{2019PNAS.Zeng9723}
Zeng, L., Jacobsen, S.~B., Sasselov, D.~D., {et~al.} 2019, Proceedings of the
  National Academy of Sciences, 116, 9723

\end{thebibliography}
\begin{appendix}
\section{Additional Figure}
\begin{figure*}[h!]
    \centering
    \caption{Posterior distributions of the best-fit DE-MCMC parameters estimating a secondary eclipse of $8 \pm 4$ ppm and  flux variations along the phase consistent with zero within 1-$\sigma$}
    \includegraphics[width=0.9\textwidth]{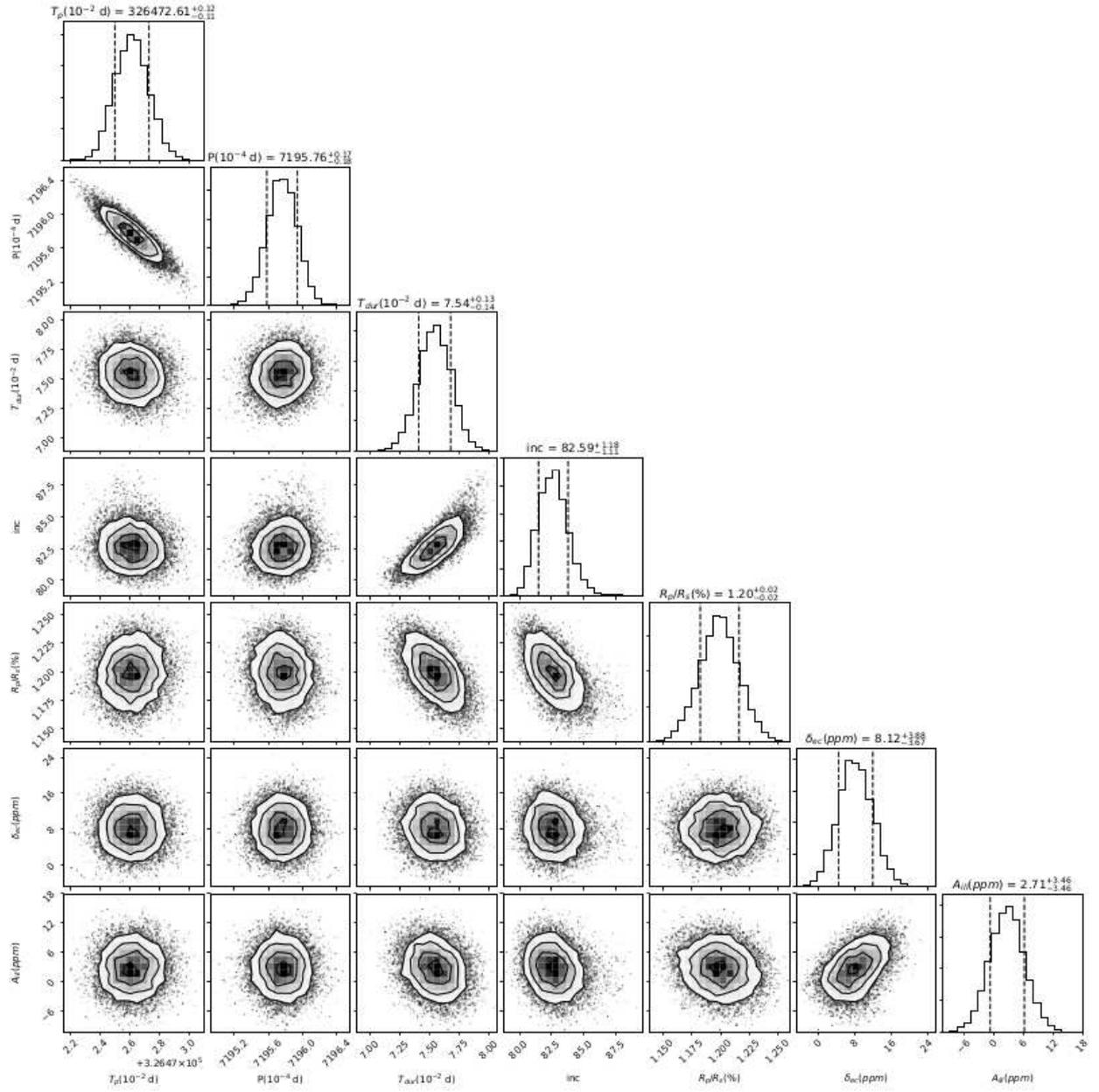}
    \label{fig:cornerPlot}
\end{figure*}

\section{Additional Table}
\begin{table*} [h!]    
\caption{Radial velocities and activity indicators obtained from HARPS-N spectra of \epk.}             
\label{table:data}      
\centering          
\begin{tabular}{r r r r r r r} \\
   \hline\hline       
\noalign{\smallskip} 
\multicolumn{1}{c}{Time} & 
\multicolumn{1}{c}{RV}& 
\multicolumn{1}{c}{$\sigma_{RV}$}& 
\multicolumn{1}{c}{FWHM}&
\multicolumn{1}{c}{BIS}&
\multicolumn{1}{c}{S$_{MW}$}&
\multicolumn{1}{c}{$\sigma_{S}$}\\
\multicolumn{1}{c}{[BJD]}& 
\multicolumn{1}{c}{[km s$^{-1}$]}& 
\multicolumn{1}{c}{[km s$^{-1}$]}&
\multicolumn{1}{c}{[km s$^{-1}$]}&
\multicolumn{1}{c}{[km s$^{-1}$]}&
\multicolumn{1}{c}{[dex]}&
\multicolumn{1}{c}{[dex]}\\
\noalign{\smallskip} 
\noalign{\smallskip} 
\hline                    
\noalign{\smallskip} 
   2458448.646247 & 8.30305 & 0.00111 & 7.95045 & 0.00861 & 0.162113 & 0.001391 \\  
   2458448.772671 & 8.30349 & 0.00119 & 7.95411 & 0.00449 & 0.163025 & 0.001516 \\
   2458449.669702 & 8.29475 & 0.00152 & 7.95168 & 0.00537 & 0.156555 & 0.002403 \\
   2458449.773103 & 8.29159 & 0.00115 & 7.94866 & 0.00412 & 0.158549 & 0.001449 \\
   2458451.628408 & 8.29713 & 0.00150 & 7.94736 & 0.00126 & 0.158349 & 0.002427 \\
   2458451.714852 & 8.29862 & 0.00110 & 7.94299 & 0.00380 & 0.155424 & 0.001332 \\
   2458451.785796 & 8.29119 & 0.00188 & 7.94585 & 0.00538 & 0.159532 & 0.003248 \\
   2458454.665720 & 8.29479 & 0.00142 & 7.94709 & -0.00305 & 0.162956 & 0.002048 \\
   2458454.704590 & 8.29715 & 0.00111 & 7.94249 & 0.00116 & 0.159342 & 0.001287 \\
   2458454.752198 & 8.29915 & 0.00136 & 7.94337 & -0.00266 & 0.159117 & 0.001873 \\
   2458454.797781 & 8.29361 & 0.00128 & 7.93432 & 0.00272 & 0.159426 & 0.001649 \\
   2458473.770783 & 8.29943 & 0.00119 & 7.95659 & 0.00664 & 0.161545 & 0.001559 \\
   2458473.781096 & 8.30081 & 0.00109 & 7.96498 & 0.00817 & 0.162527 & 0.001327 \\
   2458473.791699 & 8.29960 & 0.00117 & 7.95804 & 0.00404 & 0.162222 & 0.001527 \\
   2458474.681928 & 8.30403 & 0.00101 & 7.95237 & -0.00024 & 0.163296 & 0.001144 \\
   2458474.692669 & 8.30467 & 0.00101 & 7.95938 & 0.00511 & 0.164737 & 0.001146 \\
   2458474.703689 & 8.30255 & 0.00096 & 7.95283 & 0.00401 & 0.162947 & 0.001036 \\
   2458477.654518 & 8.30892 & 0.00211 & 7.94589 & 0.00471 & 0.172257 & 0.003895 \\
   2458477.665260 & 8.30460 & 0.00211 & 7.96164 & 0.00959 & 0.166754 & 0.003864 \\
   2458477.675932 & 8.30998 & 0.00179 & 7.95751 & 0.01254 & 0.168755 & 0.002974 \\
   2458479.708679 & 8.30153 & 0.00118 & 7.95089 & 0.00670 & 0.160444 & 0.001557 \\
   2458479.719305 & 8.30017 & 0.00111 & 7.95341 & 0.00431 & 0.159743 & 0.001382 \\
   2458480.636427 & 8.29665 & 0.00232 & 7.94991 & -0.00231 & 0.160342 & 0.004680 \\
   2458481.637212 & 8.30665 & 0.00120 & 7.94791 & 0.00239 & 0.161276 & 0.001605 \\
   2458481.699103 & 8.30997 & 0.00141 & 7.94381 & 0.00502 & 0.158630 & 0.002093 \\
   2458481.754002 & 8.30768 & 0.00102 & 7.94708 & 0.00106 & 0.161390 & 0.001187 \\
   2458481.796019 & 8.30747 & 0.00134 & 7.94384 & -0.00159 & 0.160545 & 0.001986 \\
   2458482.612461 & 8.30880 & 0.00115 & 7.95844 & 0.00316 & 0.160658 & 0.001371 \\
   2458482.654594 & 8.30853 & 0.00111 & 7.94647 & -0.00178 & 0.159934 & 0.001266 \\
   2458482.717063 & 8.30555 & 0.00129 & 7.95250 & 0.00203 & 0.157578 & 0.001682 \\
   2458483.614725 & 8.30070 & 0.00111 & 7.95467 & -0.00076 & 0.161266 & 0.001353 \\
   2458483.658652 & 8.30180 & 0.00124 & 7.96441 & 0.00195 & 0.162277 & 0.001661 \\
   2458483.719165 & 8.30054 & 0.00116 & 7.95851 & 0.00274 & 0.160862 & 0.001448 \\
   2458484.616073 & 8.30533 & 0.00136 & 7.96183 & 0.00593 & 0.161610 & 0.001950 \\
   2458484.641526 & 8.30585 & 0.00147 & 7.95466 & -0.00048 & 0.160043 & 0.002214 \\
   2458484.710292 & 8.30514 & 0.00166 & 7.95450 & 0.00756& 0.163199 & 0.002732 \\
   2458484.756938 & 8.30369 & 0.00196 & 7.95362 & 0.00108 & 0.157475 & 0.003652 \\
   2458486.686940 & 8.30034 & 0.00115 & 7.95362 & 0.00445 & 0.163930 & 0.001417 \\
   2458487.573821 & 8.30361 & 0.00144 & 7.95016 & 0.00351 & 0.161553 & 0.002173 \\
   2458487.649265 & 8.30335 & 0.00131 & 7.95302 & -0.00069 & 0.160311 & 0.001833 \\
   2458487.716804 & 8.30422 & 0.00102 & 7.94868 & -0.00168 & 0.162122 & 0.001149 \\
   2458488.580811 & 8.30495 & 0.00155 & 7.94594 & 0.00548 & 0.161432 & 0.002370 \\
   2458488.662992 & 8.30392 & 0.00124 & 7.94533 & 0.00330 & 0.161097 & 0.001609 \\
   2458488.726410 & 8.30444 & 0.00124 & 7.95802 & 0.00138 & 0.164520 & 0.001582 \\
   2458489.623601 & 8.30849 & 0.00117 & 7.95603 & -0.00245 & 0.160676 & 0.001516 \\
   2458489.682297 & 8.31146 & 0.00124 & 7.95658 & -0.00029 & 0.161087 & 0.001665 \\
   2458489.755993 & 8.30831 & 0.00206 & 7.95408 & -0.00505 & 0.165908 & 0.004034 \\
   2458491.552420 & 8.30266 & 0.00112 & 7.95039 & 0.00256 & 0.161762 & 0.001374 \\
   2458491.660412 & 8.30458 & 0.00108 & 7.95327 & 0.00153 & 0.160834 & 0.001302 \\
   2458492.553386 & 8.30806 & 0.00166 & 7.96246 & 0.00528 & 0.165291 & 0.002716 \\
   2458492.707861 & 8.30797 & 0.00138 & 7.95018 & 0.00140 & 0.165067 & 0.002005 \\
\noalign{\smallskip} 
\hline
\end{tabular}
 \end{table*}  
 \begin{table*}[h]
 \centering
\begin{tabular}{r r r r r r r} \\
   \hline\hline        
\noalign{\smallskip} 
\multicolumn{1}{c}{Time} & 
\multicolumn{1}{c}{RV}& 
\multicolumn{1}{c}{$\sigma_{RV}$}& 
\multicolumn{1}{c}{FWHM}&
\multicolumn{1}{c}{BIS}&
\multicolumn{1}{c}{S$_{MW}$}&
\multicolumn{1}{c}{$\sigma_{S}$}\\
\multicolumn{1}{c}{[BJD]}& 
\multicolumn{1}{c}{[km s$^{-1}$]}& 
\multicolumn{1}{c}{[km s$^{-1}$]}&
\multicolumn{1}{c}{[km s$^{-1}$]}&
\multicolumn{1}{c}{[km s$^{-1}$]}&
\multicolumn{1}{c}{[dex]}&
\multicolumn{1}{c}{[dex]}\\
\noalign{\smallskip} 
\hline             
\noalign{\smallskip} 
   2458495.748772 & 8.31198 & 0.00135 & 7.95525 & 0.00214 & 0.165460 & 0.001854 \\
   2458495.783912 & 8.31158 & 0.00154 & 7.96438 & -0.00013 & 0.165436 & 0.002443 \\
   2458502.601755 & 8.31440 & 0.00137 & 7.95641 & 0.00055 & 0.161180 & 0.001846 \\
   2458503.609625 & 8.30697 & 0.00091 & 7.95369 & 0.00419 & 0.161817 & 0.000932 \\
   2458503.693841 & 8.30548 & 0.00114 & 7.95433 & 0.00269 & 0.164623 & 0.001387 \\
   2458504.585941 & 8.30416 & 0.00129 & 7.95530 & 0.00767 & 0.164177 & 0.001686 \\
   2458504.682403 & 8.30667 & 0.00148 & 7.95765 & -0.00128 & 0.166852 & 0.002172 \\
   2458516.672237 & 8.30868 & 0.00128 & 7.96298 & 0.00216 & 0.167572 & 0.001761 \\
   2458517.514710 & 8.30508 & 0.00135 & 7.95908 & 0.00214 & 0.166963 & 0.001919 \\
   2458518.621166 & 8.30784 & 0.00237 & 7.96813 & 0.00395 & 0.164545 & 0.004906 \\
   2458520.635494 & 8.30980 & 0.00143 & 7.94928 & 0.00300 & 0.162982 & 0.002147 \\
   2458521.620497 & 8.31065 & 0.00098 & 7.95091 & 0.00359 & 0.163437 & 0.001057 \\
   2458522.564861 & 8.30982 & 0.00123 & 7.95467 & 0.00260 & 0.162507 & 0.001627 \\
   2458523.506226 & 8.32049 & 0.00110 & 7.95280 & -0.00021 & 0.164704 & 0.001352 \\
   2458523.694976 & 8.31865 & 0.00164 & 7.95742 & 0.00341 & 0.168972 & 0.002799 \\
   2458524.605888 & 8.31309 & 0.00113 & 7.95683 & 0.00014 & 0.164166 & 0.001406 \\
   2458525.497527 & 8.31490 & 0.00215 & 7.97070 & -0.00436 & 0.167561 & 0.004194 \\
   2458526.494454 & 8.31922 & 0.00110 & 7.95798 & 0.00642 & 0.166978 & 0.001349 \\
   2458526.661328 & 8.31420 & 0.00138 & 7.96095 & 0.00259 & 0.163577 & 0.002069 \\
   2458527.463463 & 8.31065 & 0.00127 & 7.96112 & -0.00134 & 0.164751 & 0.001770 \\
   2458527.549018 & 8.31147 & 0.00122 & 7.96269 & 0.00463 & 0.163105 & 0.001626 \\
   2458528.477458 & 8.31350 & 0.00139 & 7.95933 & 0.00345 & 0.168473 & 0.002064 \\
   2458528.556265 & 8.31546 & 0.00140 & 7.96564 & 0.00743 & 0.161408 & 0.002055 \\
   2458531.558761 & 8.31582 & 0.00134 & 7.94977 & 0.00601 & 0.166320 & 0.001928 \\
   2458531.639813 & 8.31424 & 0.00148 & 7.94502 & 0.00322 & 0.166477 & 0.002284 \\
   2458536.471164 & 8.31924 & 0.00216 & 7.95262 & 0.00684 & 0.164220 & 0.004151 \\
   2458536.656704 & 8.31464 & 0.00175 & 7.95197 & 0.00388 & 0.162231 & 0.003119 \\
   2458537.517898 & 8.30821 & 0.00146 & 7.95749 & 0.00472 & 0.166541 & 0.002242 \\
   2458537.635603 & 8.31090 & 0.00297 & 7.96045 & 0.00673 & 0.163413 & 0.006965 \\
   2458538.535197 & 8.31267 & 0.00181 & 7.94527 & 0.00395 & 0.157503 & 0.003217 \\
   2458538.648770 & 8.31338 & 0.00218 & 7.95719 & -0.00158 & 0.163335 & 0.004388 \\
   2458539.511663 & 8.31262 & 0.00104 & 7.94413 & 0.00779 & 0.159633 & 0.001198 \\
   2458539.649830 & 8.30992 & 0.00134 & 7.94932 & -0.00014 & 0.161489 & 0.001955 \\
   2458540.599790 & 8.31102 & 0.00173 & 7.94556 & 0.00610 & 0.166156 & 0.002955 \\
   2458544.518534 & 8.32141 & 0.00113 & 7.95140 & 0.00360 & 0.165684 & 0.001413 \\
   2458545.504538 & 8.31646 & 0.00128 & 7.95863 & 0.00605 & 0.164817 & 0.001778 \\
   2458546.534312 & 8.32454 & 0.00108 & 7.96089 & 0.00475 & 0.167815 & 0.001328 \\
   2458547.521841 & 8.31733 & 0.00115 & 7.96030 & -0.00106 & 0.161731 & 0.001456 \\
   2458563.485321 & 8.31693 & 0.00119 & 7.95127 & 0.00312 & 0.163542 & 0.001558 \\
   2458564.452469 & 8.32557 & 0.00197 & 7.95614 & 0.00187 & 0.162222 &  0.003554\\
   2458565.419383 & 8.32137 & 0.00113 & 7.95627 & 0.00559 & 0.159284 & 0.001386 \\
   2458565.512085 & 8.31756 & 0.00146 & 7.96220 & -0.00176 & 0.161004 & 0.002198 \\
   2458566.410589 & 8.31313 & 0.00097 & 7.96109 & 0.00492 & 0.160862 &  0.000993 \\
   2458566.585890 & 8.31922 & 0.00143 & 7.95435 & 0.00338 & 0.166439 &  0.002095 \\
   2458591.467886 & 8.32028 & 0.00148 & 7.96297 & 0.00209 & 0.166725 & 0.002231 \\
   2458592.438144 & 8.32905 & 0.00140 & 7.95954 & -0.00143 & 0.162464 & 0.002022 \\
   2458593.429973 & 8.32744 & 0.00123 & 7.95861 & 0.00622 & 0.167189 & 0.001604 \\
   2458594.431732 & 8.32238 & 0.00131 & 7.96344 & 0.00381 & 0.166443 & 0.001810 \\
   2458595.442042 & 8.32725 & 0.00114 & 7.95426 & 0.00223 & 0.165965 &  0.001369\\
   2458596.459273 & 8.32153 & 0.00141 & 7.95408 & 0.00289 & 0.165924 & 0.002010 \\
   2458597.431449 & 8.32030 & 0.00123 & 7.96288 & 0.00437 & 0.164891 & 0.001581 \\
   2458598.403845 & 8.32899 & 0.00090 & 7.95529 & 0.00116 & 0.163372 & 0.000861 \\
   2458604.418890 & 8.31734 & 0.00124 & 7.95691 & 0.00482 & 0.158740 & 0.001624 \\
   2458605.424834 & 8.32036 & 0.00104 & 7.95534 & -0.00062 & 0.160055 & 0.001140 \\
   2458606.417642 & 8.32380 & 0.00107 & 7.95613 & 0.00348 & 0.156840 & 0.001216 \\
   2458607.427230 & 8.31996 & 0.00107 & 7.95816 & -0.00308 & 0.160655 & 0.001216 \\
   2458609.400625 & 8.32538 & 0.00129 & 7.95423 & 0.00632 & 0.164608 & 0.001752 \\
\noalign{\smallskip} 
\hline                   
\end{tabular}
\end{table*}
\end{appendix}

\end{document}